\documentclass[%
 reprint,
 amsmath,amssymb,
 aps,
]{revtex4-2}

\usepackage{subcaption}
\usepackage{amsmath}
\usepackage{lmodern,babel,adjustbox,booktabs,multirow}
\usepackage[dvipsnames,table,xcdraw]{xcolor}
\usepackage{upgreek}
\usepackage{enumitem}
\usepackage{tabularx}
\usepackage{changepage} 
\usepackage{esint} 

\begin{document}

\preprint{}

\title{Role of interchange and mixed modes in magnetic Rayleigh-Taylor instability dynamics --- a 2D vs 3D study}
\author{Manohar Teja Kalluri}
\email{manohartejakalluri25@gmail.com}
\author{Andrew Hillier}
\affiliation{Department of Mathematics and Statistics, University of Exeter, Exeter, United Kingdom}

\date{\today}

\begin{abstract}
The magnetic Rayleigh–Taylor instability (MRTI) governs plasma mixing and transport in diverse systems. Owing to computational constraints, two-dimensional (2D) MRTI simulations are common. It is intuitive that 2D and three-dimensional (3D) counterparts differ significantly. One of the crucial differences is the types of wave modes admissible in the two systems. Perturbations in 3D MRTI facilitate the coexistence of interchange, undular, and mixed modes. Such rich physics is not possible in 2D, where the perturbations can only have undular modes. This raises the question, ``\textit{how do the interchange and mixed modes change the MRTI dynamics?}" To address this aspect, we simulate MRTI in 2D and 3D, across different magnetic field strengths. Consequently, this paper provides, currently lacking, a rigorous qualitative and quantitative evaluation of the similarities and differences between 2D and 3D MRTI.

The 2D MRTI is characterized by large-scale plumes with inefficient mixing and significant fluid interpenetration, while the 3D is characterized by significant mixing and small-scale plumes. Consequently, higher gravitational potential energy is extracted in 2D. The small-scale vortices and large current sheets lead to higher energy dissipation in 3D. Extending to fundamental nature of evolution, 3D simulations tend towards a self-similar state (due to interchange and mixed modes). But such behaviour was not observed in our 2D simulations, due to the persistent influence of magnetic field. Surprisingly, the reconnection driven magnetic-to-kinetic energy conversion is similar in 2D and 3D. Overall, the work concludes that interchange and mixed modes influence the dynamics significantly leading to stark differences between 2D and 3D MRTI. The work thus guides the extent to which one can extrapolate the 2D simulations results to understand MRTI in astrophysical and fusion energy systems.
\end{abstract}

\maketitle

\section{Introduction} \label{intro}

\begin{figure*}
    \centering
    \includegraphics[width = \textwidth]{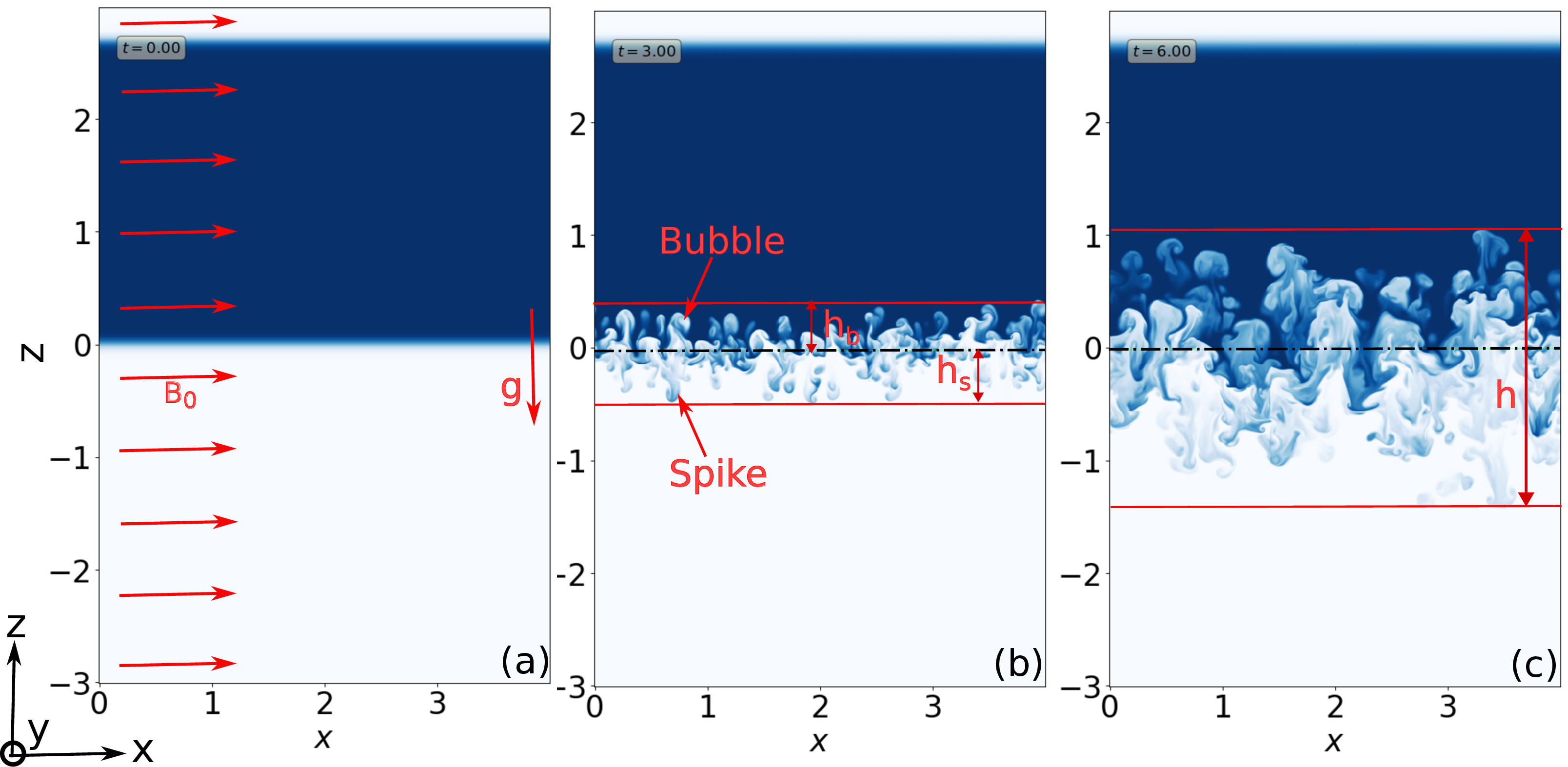}
    \caption{Figure showing the initial configuration (a), and evolution (b, c) of MRTI mixing layer through the density contours (2D slice at mid $y$-plane) at different time instants $t {=} 3.0,$ $t {=} 6.0$. The snapshots correspond to $B_0 = 5\% B_c$ case. $B_0$ is externally imposed uniform magnetic field strength. $B_c \left( {=} \sqrt{\frac{(\rho_h - \rho_l) g}{2 k cos^2 \theta}} \right)$ is the critical magnetic field strength at which the 2D instability onset is fully suppressed. The red lines mark the boundaries of the mixing layer, whose height is $h$. The dashed dotted black line is the centre line $z {=} 0$.}
    \label{MRTI_config}
\end{figure*}

When a fluid of high-density ($\rho_h$) is supported by another fluid of low-density ($\rho_l$) in the presence of gravitational and magnetic fields, perturbations at the interface of the two fluids \textit{could} lead to interpenetration of one fluid into the other. This phenomenon is called the \textit{magnetic Rayleigh-Taylor instability} (MRTI) \citep{Rayleigh_1883, Taylor1950, Kruskal1952, Roberts1963}. MRTI leads to interpenetration and mixing of fluids. This region is called the \textit{mixing layer}, and it grows over time. Figure \ref{MRTI_config} shows the temporal evolution of MRTI and the mixing layer. The dark blue and white indicate the high- and low-density regions, respectively. The intermediate densities, resulting from fluid mixing, are in light blue. The two red lines mark the boundaries of mixing layer. MRTI plays crucial role in transportation and mixing of plasma in laboratory and astrophysical systems \citep{ZHOU2017_1, ZHOU2017_2, Zhou2024}. 

MRTI in these systems involves complex physics, making it computationally demanding. 
It is at present standard practice in astrophysics to use 2D simulations to study nonlinear MRTI because computational costs are substantially lower \citep{Changmai2023, Porth2014, Zhdankin2023}, but while it has been presumed that 2D simulations are a reasonable proxy for 3D physics, there is already well-established knowledge \citep{Boffetta_2011, Zhou_2024} from the non-magnetic turbulence literature that should alert us to important differences between statistical features of 2D and 3D turbulence. Furthermore, once magnetic fields are introduced, a broader range of wave modes become available in 3D than in 2D and thus there are yet more reasons to anticipate greater differences. To elaborate on the different wave modes, we start with the linear MRTI theory. The linear growth rate for an MRTI with single mode plane wave perturbation is given by \citep{Chandrasekhar_1961}
\begin{equation}
    \sigma_{mhd} = \sqrt{\frac{\rho_h {-} \rho_l}{\rho_h {+} \rho_l} k g {-} \frac{2 k^2 B^2 cos^2 \theta}{(\rho_h {+} \rho_l)}}.
    \label{Growthrate_MRT}
\end{equation}
Here $k$ is the perturbation wave mode, $B$ is the magnetic field strength, and $\theta$ is the angle between $\textbf{k}$ and $\textbf{B}$. Bold represents the vector quantity and corresponding non-bold represents their magnitude. Here, \textbf{B} is the total magnetic field, that is sum of initial ($\mathbf{B_0}$) and perturbed ($\mathbf{b}$) magnetic fields. We find that, for a given $\rho_h, \rho_l, g, k$, and $B$ growth rate is least when $\theta$ is 0 ($\textbf{k} \parallel \textbf{B}$, called undular modes), and largest when $\theta = \pi/2$ ($\textbf{k} \perp \textbf{B}$, called interchange modes). Modes at a different angle, $0 < \theta < \pi/2$, are called mixed modes.
\begin{figure*}
    \centering
    \includegraphics[width=\linewidth]{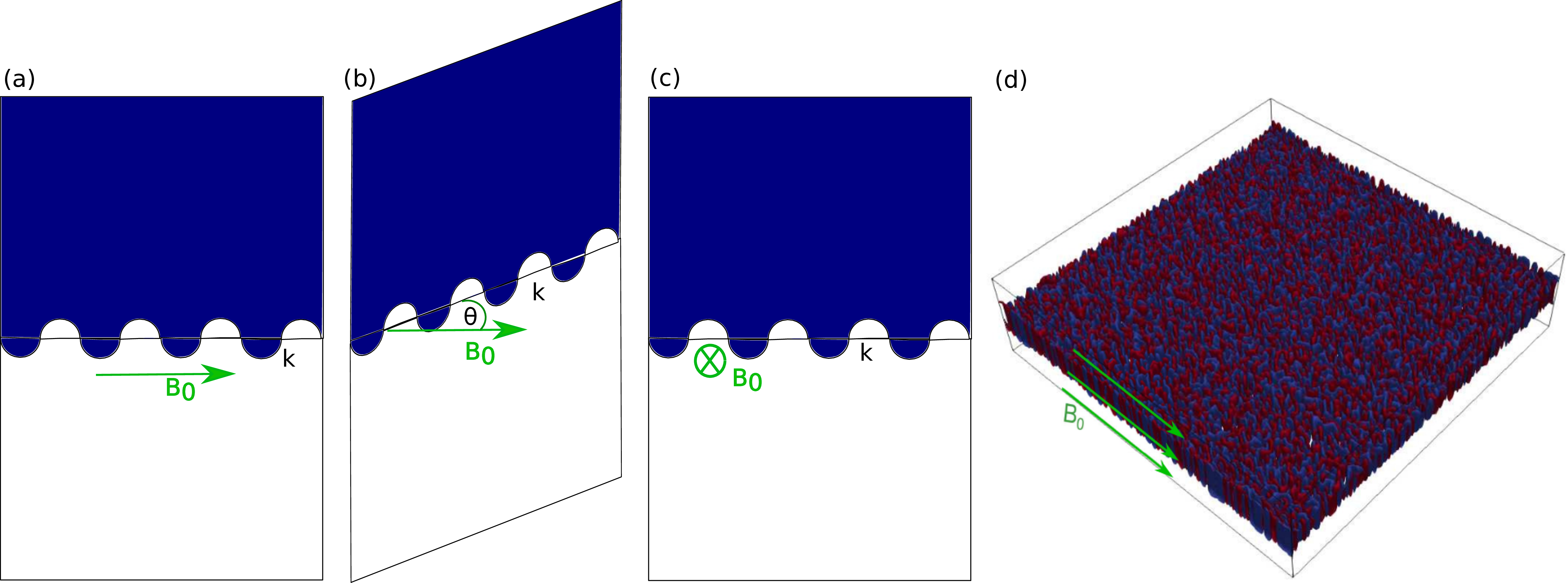}
    \caption{Schematic figure showing a different wave modes: (a) undular modes, perturbations parallel to imposed magnetic field $B_0$ (current 2D MRTI configuration); (b) mixed modes, perturbations inclined at non-perpendicular angle with respect to $B_0$; (a) interchange modes, perturbations perpendicular to $B_0$. \textit{(d)} Instantaneous velocity contour of a 3D MRTI at time $t = 0.2$. Red (blue) corresponds to positive (negative) velocity.}
    \label{pert}
\end{figure*}

In a 2D system, the interface is a line with the magnetic field imposed at an angle to the perturbation wave modes. Thus, the perturbation has either undular (as shown in figure \ref{pert}(a)) or interchange (figure \ref{pert}(c)) or mixed (figure \ref{pert}(b)) modes. In contrast, the interface in a 3D system is a plane ($xy$-plane). While the magnetic field remains directed along $x$, perturbation wave modes can exist anywhere within the $xy$-plane, as shown in figure \ref{pert}\textit{(d)}. This leads to a broad spectrum of wave modes --- undular ($k_x \neq 0, k_y = 0$), interchange (modes with $k_x = 0, k_y \neq 0$), mixed modes ($0 {<} \theta {<} \pi/2$, $k_x {\neq} 0, k_y {\neq} 0$), and coupled modes coming from product of $k_x$ and $k_y$ modes (see equation \ref{white_noise_3d} for example). As a result, 3D MRTI exhibits a rich variety of wave behaviours, producing complex dynamics arising from these different mode types. The evolution of these different modes from early stages of instability were shown in the figure \ref{pert} (d). On the contrary, the 2D MRTI dynamics are governed exclusively by modes at a fixed angle with respect to the magnetic field, as shown in figure \ref{pert}\textit{(a,b,c)}. While the differences are expected to influence the dynamics in the two systems, it is not well understood how the additional interchange and mixed modes change the MRTI dynamics. 

Besides, a rigorous qualitative and quantitative evaluation of the similarities and differences between 2D and 3D MRTI is lacking. While \citet{Jun1995} examined MRTI in 2D and 3D, their analysis was limited to comparing kinetic and magnetic energies. Several other important aspects, such as fluid mixing, energy dissipation, energy anisotropy, and non-linear growth rates, remain unexplored. Further, magnetic field strength is known to significantly influence RTI behaviour, like suppressing small-scale wave modes and dissipation, increasing anisotropy and non-linear growth \citep{Stone2007a, kalluri2025_self-similarity}. But, \citet{Jun1995} investigated at a single magnetic field strength, $B_0 = 2\%B_c$ (for $B_c$ see caption of figure \ref{MRTI_config}). These findings suggest that conclusions drawn from comparing 2D and 3D MRTI at a single field strength may not hold across different magnetic field strengths.

Through systematic analysis, this work addresses: \\
\textit{(i)} How do interchange and mixed modes change the MRTI dynamics? \\
\textit{(ii)} How does 2D and 3D MRTI dynamics compare qualitatively and quantitatively? \\
\textit{(iii)} How does the above difference change with magnetic field strength? \\
We provide a comprehensive comparison of the mixing layer characteristics and energy dynamics across a range of magnetic field strengths, from $B_0 {=} 1\% B_c$ to $15\% B_c$. 
\vspace{-5pt}
\section{Numerical Methodology} \label{Methodology}
In the present study, MRTI is modelled numerically by superimposing a high density fluid $(\rho_h)$ over a low density fluid $(\rho_l)$ in the presence of an externally imposed uniform, unidirectional magnetic $(\mathbf{B_0})$ and gravitational $(\mathbf{g})$ fields. $\mathbf{B_0}$ and $\mathbf{g}$ are along the horizontal ($x$) and vertical ($z$) directions, respectively. The numerical modelling was performed using Dedalus \citep{Dedalus2020}, an open-source, parallelized computational framework to solve the partial differential equations using the spectral method. We solve the non-ideal MHD governing equations \citep{kalluri2025_self-similarity}
\begin{subequations}
    \begin{align}
        \partial_t \mathbf{u} {-} \nu \mathbf{\nabla}^2 \mathbf{u} = {-} (\mathbf{u} {\cdot} \mathbf{\nabla}) \mathbf{u} & {-} \frac{1}{\rho} \mathbf{\nabla} p {-} \frac{\delta \rho}{\rho}\mathbf{g} {-} \frac{1}{\rho} (\mathbf{B} {\cdot}  \mathbf{\nabla}) \mathbf{B}, \\
        \partial_t \mathbf{B} {-} \eta \mathbf{\nabla}^2 \mathbf{B} {-} c_p^2 \mathbf{\nabla}(\mathbf{\nabla} {\cdot} \mathbf{B}) & = (\mathbf{B} {\cdot} \mathbf{\nabla}) \mathbf{u} {-} (\mathbf{u} {\cdot} \mathbf{\nabla}) \mathbf{B}, \\
        \partial_t \rho {-} \mathcal{D} \mathbf{\nabla}^2 \rho & = {-} (\mathbf{u} {\cdot} \mathbf{\nabla}) \rho,  \\
        \mathbf{\nabla} {\cdot}  \mathbf{u} & = 0.
    \end{align}
    \label{MHDeqns_1}
\end{subequations}
\hspace{-10pt} $\textbf{u}$ and $\rho$ represent the velocity and density. $\delta \rho$ represent $(\rho - \rho_0)$, $\rho_0$ being the initial density profile. The solenoidal condition for the magnetic field ($\mathbf{\nabla} \cdot \mathbf{B} = 0$) is ensured through divergence cleaning term ($c_p^2 \mathbf{\nabla}(\mathbf{\nabla} \cdot \mathbf{B})$) \citep{DEDNER_2002}. In the present study, the value of $c_p$ is set such that the solenoid condition is satisfied to machine precision throughout the simulation. We consider two miscible plasmas with equal and constant density diffusion coefficient ($\mathcal{D}$) \citep{Briard_Gréa_Nguyen_2024}. $\nu, \eta$ are the coefficients of fluid viscosity and magnetic diffusion, respectively. $\nu, \eta,$ and $\mathcal{D}$ is set to $10^{-4}$. 

The velocity divergence for the hydrodynamic (HD) variable density systems (like RTI) is typically non-zero and of the form, $\mathbf{\nabla} {\cdot}  \mathbf{u} {=} {-}\frac{1}{Re Sc} \nabla {\cdot} \left(\frac{1}{\rho} \nabla\rho \right)$ \citep{Sandoval1995}. However, in natural systems the density gradients are typically several orders of magnitude smaller than the $Re Sc$ (i.e., $\nabla \rho {\ll} ReSc$), leading to $\nabla \cdot \mathbf{u} {\approx} 0$ \citep{Cabot_2013}. Hence, in this study, we solve an approximation of the Variable Density problem, with $\nabla \cdot \mathbf{u} {=} 0$. The sensitivity of RTI dynamics to velocity divergence criteria meant that it is possible that the approximation could lead to quantitative differences in small-scale mixing, non-linear growth constant, turbulent kinetic and magnetic energies, and energy dissipation. 

The present MRTI study uses periodic boundary conditions in all directions. In 2D, the domain is $L_x {\times} L_z {=}$ $ 4 {\times} 6$ units with a resolution of $2048 {\times} 3072$, and in 3D, the domain is $L_x {\times} L_y {\times} L_z {=}$ $ 4 {\times} 4 {\times} 6$ units with resolution $512 {\times} 512 {\times} 768$. In both, $x{\in}[0, L_x]$, $y{\in}[0, L_y]$, $z{\in}[-L_z/2, L_z/2]$. $g$ is taken as 1. $\rho_h {=} 3$ and $\rho_l {=} 1$ for all the simulations. We present the results of three magnetic field strengths --- a weak field case ($B_0 = 1\%B_c$) where the mixing layer is highly turbulent similar to the HD RTI case, an intermediate field case ($B_0 = 5\%B_c$) where the mixing layer has transits from HD to magnetic field dominant, and a relatively stronger field case ($B_0 = 15\%B_c$), where the dynamics are relatively dominated by magnetic field.

The initial density profile is given by, 
\begin{equation}
    \rho = 1 {-} \frac{\Delta \rho}{2} \left[ \tanh{\left(\frac{z {-} 0.45 L_z}{0.05} \right)} {+} 1 \right] {+} \frac{\Delta \rho}{2} \left[ \tanh{\left(\frac{z}{0.05} \right)} {+} 1 \right] .
    \label{rho_profile}
\end{equation}
The above density profile has two interfaces --- one at $z {=} 0$, other at $z {=} 0.45 L_z$ (see figure \ref{MRTI_config}). This facilitates periodicity in the $z$-direction. As the MRTI evolves, the interface elongates and thins (due to magnetic flux conservation) over time. Throughout the paper, ``time" refers to raw dimensionless simulation time. While the density diffusion aids in smoothing the density profile, it is important to ensure that the interface is initially resolved over an adequate number of grid points, so that there are a sufficient number of grid points at any time during evolution. Hence, we chose a hyperbolic tangent profile with a half-width of 0.05 (${\approx} 7$ grid points) so that the interface remains adequately resolved in 3D. For consistency, we use the same 0.05 half-width (${\approx} 25$ grid points) in 2D. 

\begin{figure}
    \centering
    \includegraphics[width=0.7\linewidth]{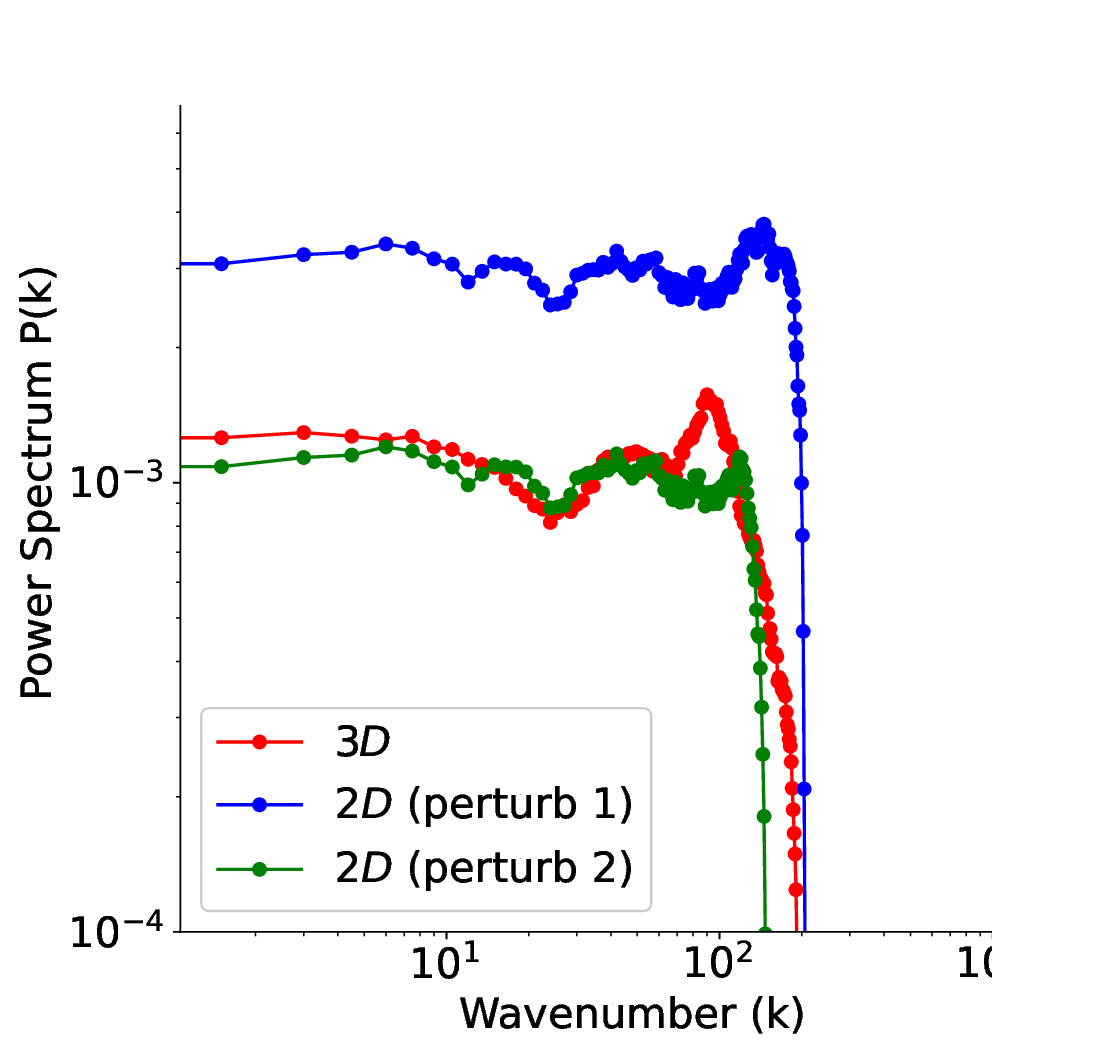}
    \caption{Power spectrum of 2D (blue) and 3D (red) perturbations. Simulations with perturb 1 spectrum (blue) are the main simulations used for analysis through out the paper. Simulation with perturb 2 is a test simulation to evidence perturb 2 and perturb 1 show similar results when quantities are compared in terms of mixing layer height.}
    \label{perturbpower}
\end{figure}

The upper stable interface is left unperturbed and it acts as a marker for boundary influence \citep{kalluri2025_self-similarity}. The lower interface $(z =  0)$ is perturbed by a vertical velocity. For 3D MRTI, the perturbation is of the form
    \begin{equation*}
    u_z = \Big[ \sum^{64}_{i=0} a_i \times \sin \left( \frac{2 \pi k_i x}{L_x} {+} \phi_i \right) \times \sum^{64}_{j=0} \sin \left( \frac{2 \pi k_j y}{L_y} {+} \phi_j \right) 
\end{equation*}
\begin{equation}
    - (a_0 \sin \phi_0 \times a_0 \sin \phi_0) \Big] e^{{-}(z^2/0.01)},
    \label{white_noise_3d}
\end{equation}
where $ a_i \in  [-10^{-3}, 10^{-3}], \phi_i \in [0, \pi]$ are randomly chosen. $k_i$ and $k_j$ take the values of $i$ and $j$ respectively to ensure we introduce all wave modes in the range ($0-64$). Thus, in 3D the system has undular $(k_i \neq 0, k_j = 0)$, interchange $(k_i = 0, k_j \neq 0)$, and a wide range of modes in between $(k_i \neq 0, k_j \neq 0)$. The perturbation for the 2D MRTI case is
\begin{equation}
    u_z {=} \sum^{128}_{k_i=1} a_i \sin \left( \frac{2 \pi k_i x}{L_x} {+} \phi_i \right) e^{{-}(z^2/0.01)},
    \label{white_noise_2d}
\end{equation}
where $ a_i {\in}  [-0.0125, 0.0125], \phi_i {\in} [0, \pi]$. The 2D case has purely undular modes so evolves less rapidly than a 3D case that admits interchange modes. We choose therefore to match perturbation power spectra, with a larger range of perturbation wave modes in 2D ($k_{max} = 128$) and those with higher amplitude compared to 3D ($k_{max} = 64$). While it might now appear that the two cases are not directly comparable, but since we study the dynamics with
respect to mixing layer height, rather than time, they become dynamically comparable. To provide evidence for this, we run a simulation case with $k_{max}^{2D} = k_{max}^{3D} \left(\equiv\sqrt{(k_x)_{max}^2 + (k_y)_{max}^2} \right) = 90$ and with smaller perturbation amplitude such that the energy spectrum of 2D (green line) and 3D are comparable. This will be discussed and evidence is provided in $\S$\ref{sec:perturbcomp}. Figure \ref{perturbpower} shows the power spectrum of the initial velocity perturbation ($u_z(x, z {=} 0, t {=} 0)$). The magnitude of the power spectrum $P(k)$ is similar in perturbed wave mode range, and falls to machine precision $O(10^{-17})$ for large wave numbers ($k {>} 10^2$). The magnetic field and perturbations are both in the $x$-direction. In 2D, the system only has undular modes. In both 2D and 3D, we introduce a wide range of wave modes ($k \in [1, 128]$ for 2D and $k \in [1, 64]$ for 3D) as suggested by \citet{ramaprabhu_2005, dalziel_1999, Dimonte2004, GLIMM2001}. The perturbations decay in a Gaussian profile about the interface $(z = 0)$.

\section{Results}

\begin{figure*}
    \centering
    \includegraphics[width=\linewidth]{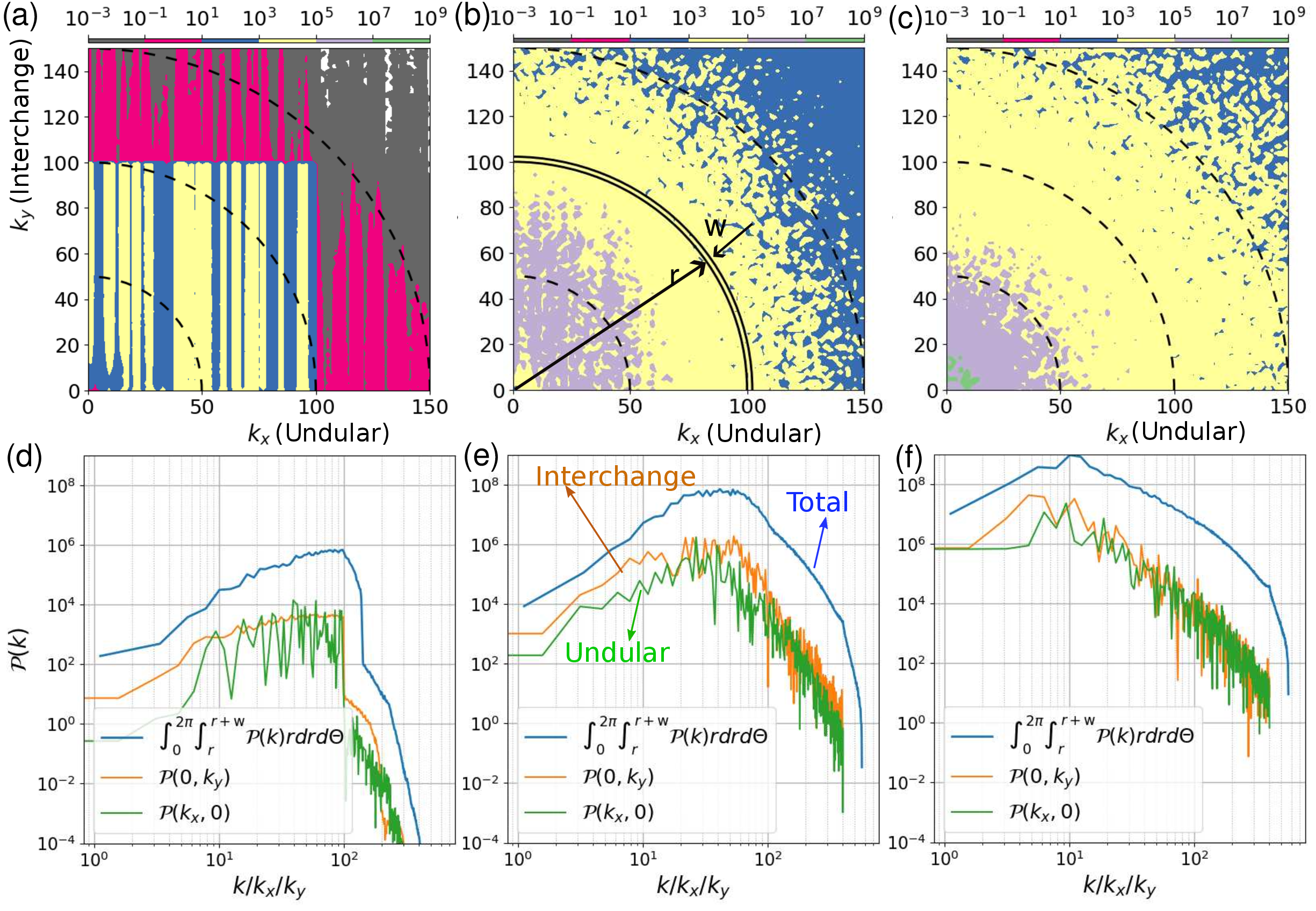}
    \caption{\textit{(top)} First quadrant of the 2D power spectrum contour. The $x$-axis represents the pure undular modes, $y$-axis represents pure interchange modes, and the region between the axes represents mixed modes. The dashed lines traces points of different $k_x, k_y$ but constant magnitude $\sqrt{k_x^2 + k_y^2}$. Each colour represents a range of energy as shown in the colour bar. \textit{(bottom)} Power spectrum of interchange (orange), undular (green) modes, and all modes (blue, obtained by radial binning into 254 bins and summing the energy in across all $k_x$ and $k_y$ in each bin), plotted at different time instants: $t = 0.2$ \textit{(left)}; $t = 2$ \textit{(centre)}; $t = 6$ \textit{(right)}. The data used to plot the energy spectrum is vertical velocity at the mid-$z$ plane, corresponding to magnetic field strength of $B_0 = 5\%B_c$.}
    \label{Pk}
\end{figure*}

The primary goal of comparing the 2D and 3D MRTI is based on the argument that 3D has a wide range of modes. This is demonstrated in the top panel of figure \ref{Pk} through 2D power spectrum contour in $k_x$ and $k_y$ space. The 2D power spectrum is calculated by taking 2D FFT of the instantaneous and vertical velocity ($u_z$) data at the mid-plane of the mixing layer ($z = 0$). The 2D FFT data is then multiplied with it's complex conjugate and the resultant data is plotted in figure \ref{Pk} (top). Note that we only plot the first quadrant since the other quadrants are mirror symmetries. The $X$-axis $(k_x \neq 0, k_y = 0)$ represents the undular mode, $Y$-axis $(k_x = 0, k_y \neq 0)$ represents the interchange mode. Each colour represents a fixed range of power, see colour bar. 

The contour plots exhibits several interesting features. Firstly, the energy contained across all wave modes increases over time, evident from the higher power purple region $(10^5-10^7)$ predominantly replacing the lower power blue and yellow regions $(10^1-10^5)$. At the late stages ($t = 6$, figure \ref{Pk}(c)), we see that the small modes $(k < 10)$ have a peak power of order $10^9$. The other important observation is the increasing power asymmetry. The dashed lines in the figure \ref{Pk} top panel shows waves modes $(k_x, k_y)$ of constant magnitude, $k \left(= \sqrt{k_x^2 + k_y^2} \right)$. We can see that in the non-linear regime, more power gets accumulated in the interchange $(k_y)$ modes --- purple colour extends to $k_y \approx 80$, but only upto $k_x \approx 50$. This makes wave modes $40 \lessapprox k \lessapprox 80$ most energetic (evident from figure \ref{Pk}(e)). Similarly, the yellow region in figure \ref{Pk}(b) (figure \ref{Pk}(c)) is restricted to $k_x \approx 140$ $(k_x \approx 150)$, but extend to $k_y \approx 160$ ($k_y \approx 200$). Because interchange modes maintains higher TKE across a wider range, they mix fluid more efficiently than undular modes.

In the bottom panel we plot the energy in interchange modes (orange line), undular modes (green line) and the total energy (blue line). The power spectrum of undular and interchange modes is plotted from the data along the $X$- and $Y$-axes of the 2D power spectrum. The total power is calculated using radial binning, i.e., considering a circular patch of width w (from radius $r = \sqrt{k_x^2+k_y^2}$ to radius $r+$w), as shown by black lines in figure \ref{Pk}(b). The total power is the sum of of power in the complete circular patch. We attribute this power to the wave mode $k \left( {=} \frac{r + (r + \text{w})}{2} \right)$. The total power spectrum was calculated at 254 such patches between 0 and $k_{max}$.

We plot the contour and spectrum at three different time instants $t = 0.2, 2, 6$. In figures \ref{Pk}(d, e, f) we see that the interchange modes have slightly higher energy compared to the undular modes. This is due to the suppression of the undular modes by the magnetic field. More importantly, we see that the energy in the undular and interchange modes together is only a small part of the total energy (blue line), which is two orders of magnitude higher than interchange or undular modes. This energy difference is in the mixed modes, implying that the mixed modes play a crucial role in the energy dynamics. This significant energy contained by mixed modes hints that physics is predominantly three-dimensional structures and not from two-dimensional structures. These crucial mixed modes cannot exist in 2D, providing the first evidence that the ability of 2D simulations to replicate 3D physics is limited. 

Following the discussion on different wave modes present in 2D and 3D MRTI, we will now delve into the how the system with interchange and mixed modes (3D) is different from the system without these (2D), qualitatively and quantitatively.

\vspace{-10pt}
\subsection{Mixing layer in 2D and 3D} 
\begin{figure*}[htbp]
    \centering
    \includegraphics[width= \linewidth]{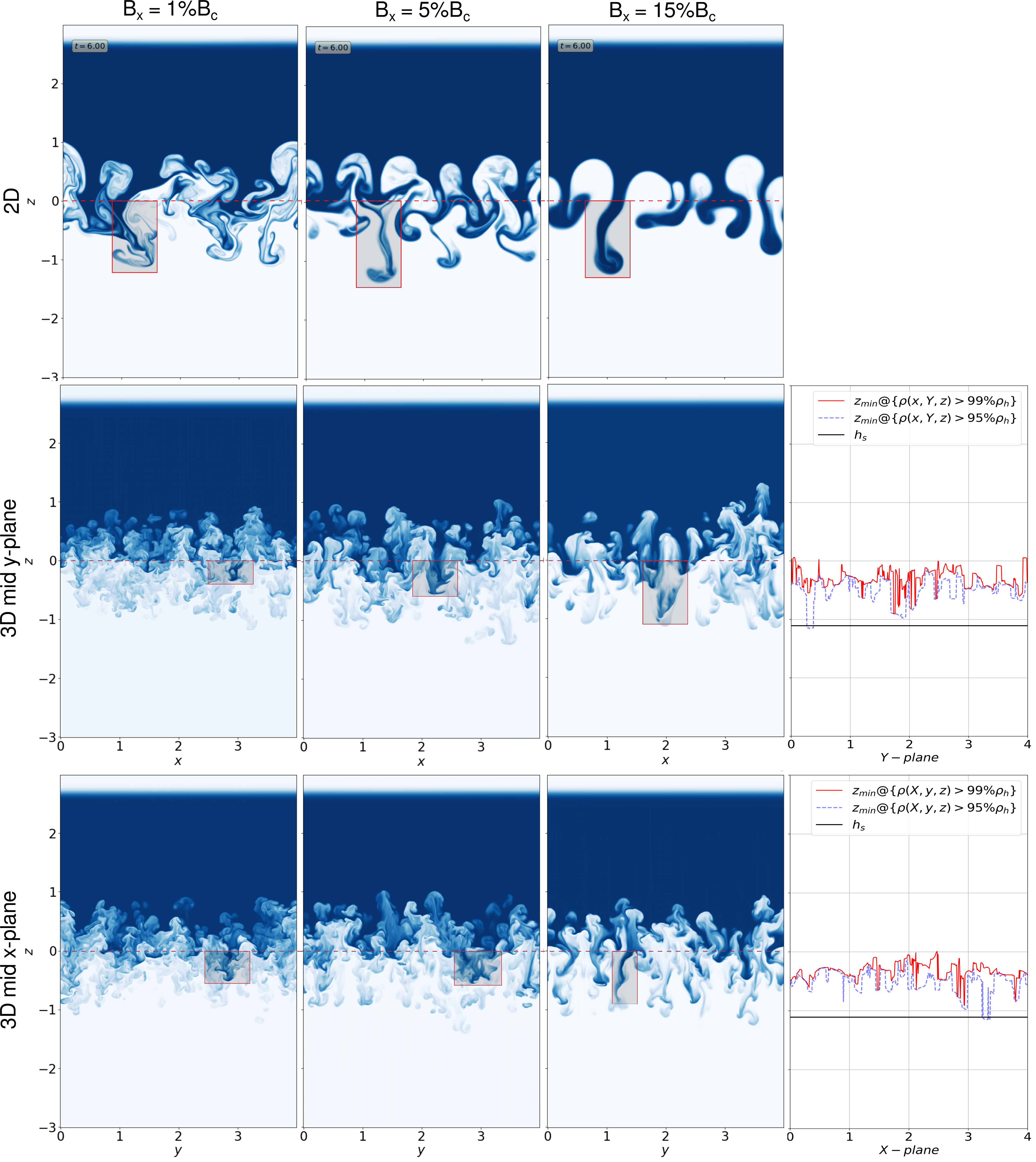}
    \caption{Comparison of mixing layer between 2D MRTI (top panel), mid-$y$ plane of 3D MRTI (middle panel), and mid-$x$ plane of 3D MRTI (bottom panel) through density contours at time $t = 6$. The left, centre and right columns show density contours at magnetic field strength $B_0 = 1\% B_c, 5\% B_c, 15\% B_c$ (right).}
    \label{2d_3d}
\end{figure*}

Starting with the mixing layer, we focus on three phenomena --- interpenetration, mixing, and stirring. Mixing is the phenomenon of blending of the high and low density fluids to yield an intermediate density fluid. If the fluid is well mixed, the final density is the arithmetic mean of two fluids. Interpenetration is the movement of fluid through the other fluid with marginal or no change in the density of fluid penetrating. Stirring is the process of mechanical deformation of flow structures through a driver (flow velocity in the present case). 

Starting with a qualitative inspection, in Figure \ref{2d_3d} we present the instantaneous density contours of 2D and 3D (mid-$y$, mid-$x$ plane slices) MRTI at different magnetic field strengths. The density contours of the 2D MRTI case (top row of figure \ref{2d_3d}) show that the plume structures are typically large-scale with the plume core largely unmixed. The fluids penetrate well into each other (highlighted with boxes) thereby, the fraction of pure fluid at the boundaries of the mixing layer is much higher in 2D. That is, 2D MRTI has higher interpenetration. 

The density contours of the 3D MRTI show that the mixing layer is made of small-scale structures and are more diffuse. The plumes with pure fluid usually do not penetrate all the way to the boundaries of the mixing layer. This is shown qualitatively from the mid-plane density contours (see the highlighted box regions) in the middle and bottom rows of figure \ref{2d_3d}. In the fourth column, we mark the vertical height up to which fluid of density $\rho > 99\% \rho_h$ (solid red line) and  $\rho > 95\% \rho_h$ (dashed blue line) penetrates at each X-, Y- plane. This demonstrates that the interpenetration of one fluid into the other is lower in 3D at all planes.

Unlike 2D, in 3D, the plumes at the boundaries of the mixing layer are finger-like and diffused. The core of the plumes is relatively well-mixed in 3D. We also see that the fraction of pure low-density fluid in the upper half of the domain is much higher in 2D compared to 3D. That is, the fluids appear to interpenetrate more in 2D compared to 3D. In both 2D and 3D, as the strength of the magnetic field increases (moving from left to right), the plumes are large scaled, less mixed, with increased interpenetration of one fluid into the other.

\subsubsection{Mixing and interpenetration} \label{sec:Mixing}

\begin{figure*}[t]
    \centering
    \includegraphics[width= 0.66\linewidth]{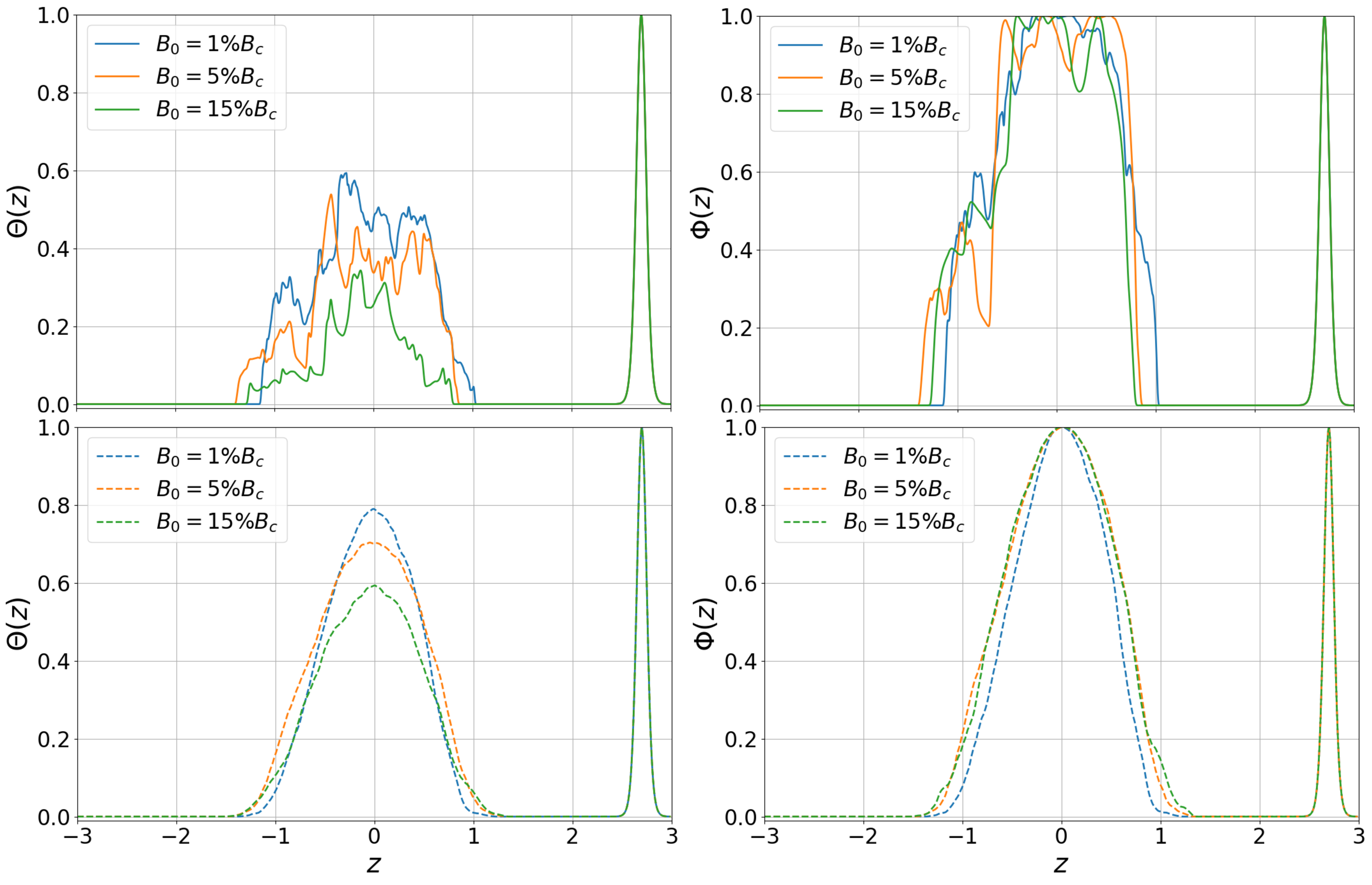}
    \caption{Comparing the spatial variation of mixing parameter (left) and dispersion parameter (right) at a given time $t = 6$, for 2D MRTI (top panel) and 3D MRTI (bottom panel) for different magnetic field strengths $B_0 = 1\% B_c, 5\% B_c, 15\% B_c$.}
    \label{thetaphi}
\end{figure*}
To quantify the degree of fluid mixing, we use \textit{mixing parameter} $\Theta$, as defined by \citet{Stone2007a}
\begin{equation}
    \Theta = 4 \left \langle \left(\frac{ \rho  - \rho_l}{\rho_h - \rho_l} \right ) \left ( \frac{\rho_h - \rho}{\rho_h - \rho_l}  \right ) \right \rangle_h,
    \label{mixing_parameter}
\end{equation}
where $\langle \star \rangle_h$ represents averaging of $\star$ along the homogeneous directions ($x$ in 2D and $x$, $y$ in 3D)
\begin{equation*}
    \langle \star \rangle_h = \frac{1}{L_x}\int_0^{L_x} \star \mathrm{d}x \text{ (2D); } \frac{1}{L_x L_y} \int_0^{L_x} \int_{-L_y/2}^{L_y/2} \star \mathrm{d}x \mathrm{d}y \text{ (3D)}.
\end{equation*}
Hence, $\Theta$ is only a function of the non-homogeneous direction. The quantities $\left( \frac{ \rho  - \rho_l}{\rho_h - \rho_l} \right)$ and $\left( \frac{ \rho_h  - \rho}{\rho_h - \rho_l} \right)$ represent the volume fraction of high and low density fluid, respectively. Outside the mixing layer, where $\rho {=} \rho_h$ or $\rho_l$, $\Theta {=} 0$. In the mixing layer, $\rho$ lies between $\rho_h$ and $\rho_l$ and $\Theta {>} 0$. When the fluid is well mixed $\rho {=} \left( \frac{\rho_h {+} \rho_l}{2} \right)$, $\Theta {=} 1$. 

Plotting the spatial variation of the mixing parameter at time $t = 6$ for different magnetic field strengths, see figure \ref{thetaphi}, we find that the mixing parameter is higher in the 3D case. The peak $\Theta$ for the $1\%B_c$ 3D MRTI case is $\approx 0.8$ (figure \ref{thetaphi}(c)), while it is $\approx 0.6$ for $1\%B_c$ 2D MRTI (figure \ref{thetaphi}(a)). The trend is similar for other magnetic field strengths as well. The 3D and 2D $5\%B_c$ MRTI cases has $\Theta \approx 0.6$ and $\Theta \approx 0.4$ respectively.

Besides the peak value of $\Theta$, the profile is very different in the two cases. In 2D, the mixing parameter has a steep jump at the boundaries of the mixing layer. This is due to the large-scale structures and a significant fraction of the pure fluid at the edge of the mixing layer. However, in the 3D case, the profile of $\Theta$ is smooth, since the fluid at the boundaries is more diffused. 

The above quantity, however, is unable to explain the interpenetration of the two fluids. To explain this, we defined the parameter $\Phi$,
\begin{equation}
    \Phi = 4 \left(\frac{ \langle \rho \rangle_h  - \rho_l}{\rho_h - \rho_l} \right ) \left ( \frac{\rho_h - \langle \rho \rangle_h}{\rho_h - \rho_l}  \right ),
    \label{interdisperse}
\end{equation}
where we average the density along the homogeneous directions and then subtract the averaged density from the low and high densities. When the two fluids are well dispersed, the homogeneous average becomes $\left( \frac{\rho_h + \rho_l}{2} \right)$. Note that for $\langle \rho \rangle_h = \frac{\rho_h + \rho_l}{2}$, the  sufficient condition is equal proportional of $\rho_h$ and $\rho_l$ at a given $z$. In such a case, irrespective of the fluid mixing, the parameter $\Phi$ attains 1 as long as the plumes penetrate well into each other. Thus, $\Phi$ informs us of the interpenetration of the two fluids.  Plotting $\Phi$, we find that $\Phi$ is 1 for most of the mixing layer region in 2D MRTI (figure \ref{thetaphi}(b)). On the contrary, $\Phi = 1$ only at the mid $z$-plane for the 3D MRTI (figure \ref{thetaphi}(d)). Thus, the two fluids interpenetrate in 2D, but are less mixed. Contrarily, in 3D, two fluids penetrate less, but mix more. The presence of interchange and mixed modes in 3D allows coupling of $k_x$ and $k_y$ modes which enhances the fluid mixing in 3D. However, the exclusive contribution of interchange and mixed modes (in 3D simulations) towards enhanced fluid mixing cannot be segregated quantitatively with the present experiments.

\subsubsection{Stirring}
Besides interpenetration and mixing, another important feature is the stirring, which refers to the mechanical deformation of fluid parcels and stretching, which increases interfacial area without necessarily homogenizing the density field. This is typically quantified based on density gradient norm $(\partial_i \rho)^2$ \citep{Emmanuel2019}.

To understand the stirring due to the flow (the mechanical driver), we derive an equation for the density gradient norm, starting from the density equation
\begin{equation}
    \partial_t \rho + u_j \partial_j \rho = \mathcal{D} \partial_{jj} \rho 
\end{equation}
Taking spatial gradient on both sides, we get
\begin{equation}
     \partial_t (\partial_i \rho) + \partial_i u_j \partial_j \rho + u_j \partial_{ij} \rho = \mathcal{D} \partial_{jj} (\partial_i \rho )
\end{equation}
Taking the dot product of above equation with $(\partial_i \rho)$ and volume averaging, after a few steps, we get
\begin{multline}
    \left \langle \frac{D}{Dt} (\partial_i \rho)^2 \right \rangle_V = - \left \langle 2 (\partial_i \rho \partial_j \rho) \partial_i u_j \right \rangle_V \\ + \left \langle \mathcal{D} \left[ \partial_{jj} (\partial_i \rho )^2 - 2(\partial_{ij} \rho)^2 \right] \right \rangle_V,
\end{multline}
where $\langle \star \rangle_V = \frac{1}{V} \int_V \star dV$ ($V$ is domain volume). 

\begin{figure*}[t]
    \centering
    \begin{subfigure}[b]{0.33\textwidth}
         \centering
        \includegraphics[width = \linewidth]{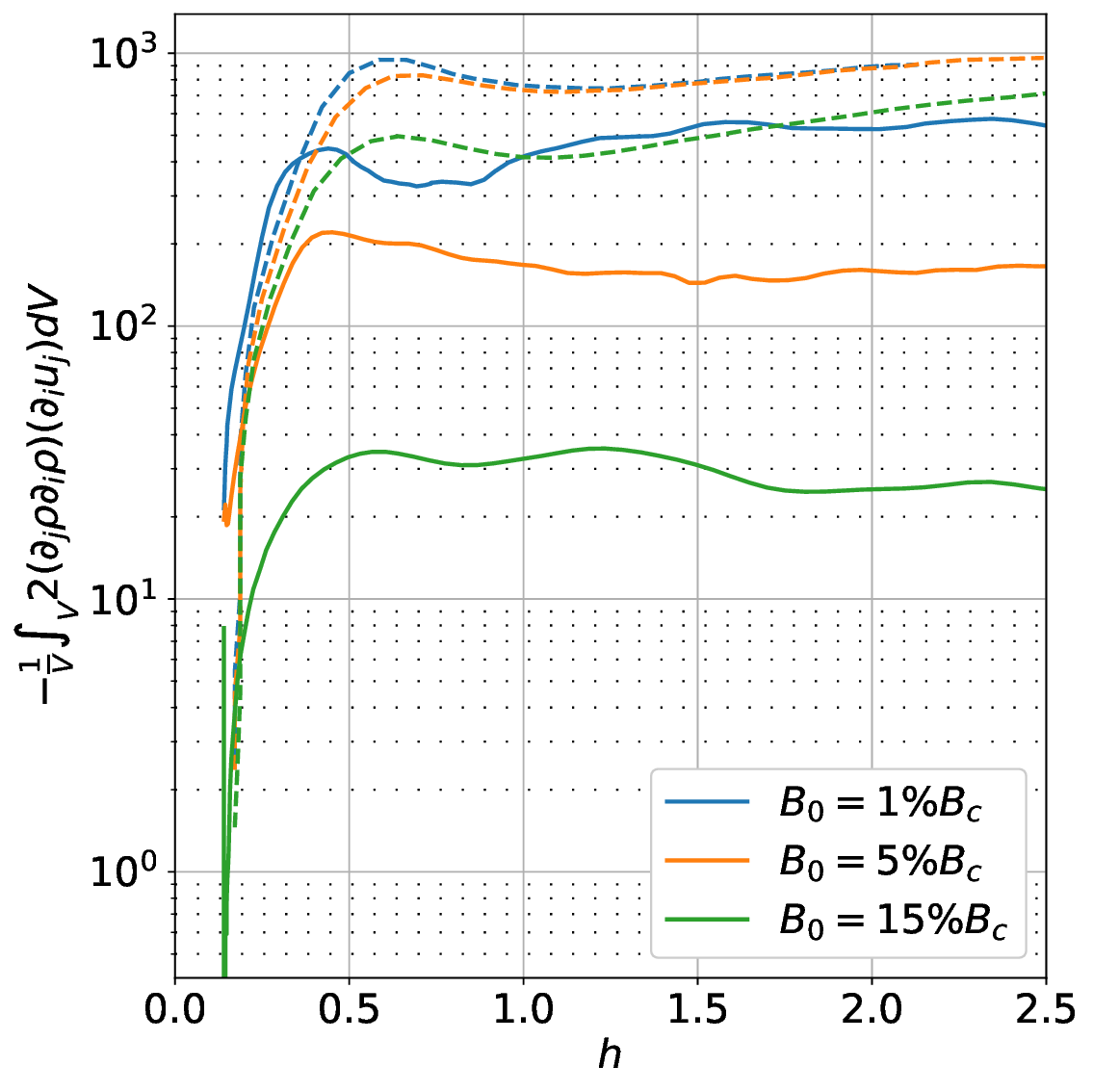}
    \end{subfigure}
    \begin{subfigure}[b]{0.33\textwidth}
         \centering
        \includegraphics[width = \linewidth]{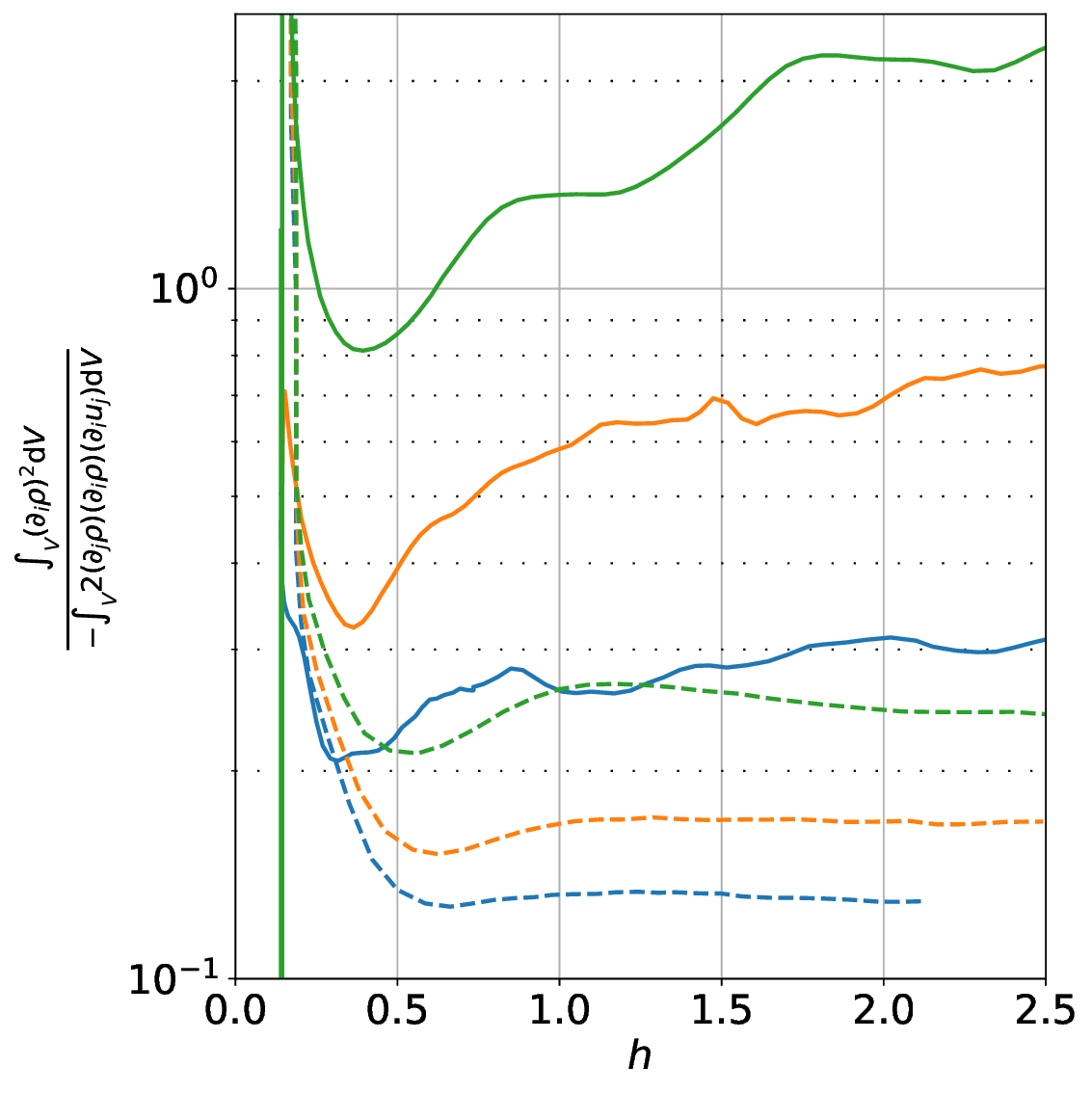}
    \end{subfigure}
    \caption{(left) Variation of: (left) $-\left \langle 2 (\partial_i \rho \partial_j \rho) \partial_i u_j \right \rangle_V $; and (right) stirring time scale $\left \langle (\partial_i \rho)^2 \right \rangle_V/-\left \langle 2(\partial_j \rho)(\partial_i \rho)(\partial_i u_j) \right \rangle_V$ with mixing layer height for 2D (solid lines) and 3D (dashed lines) MRTI, for different magnetic field strengths, $B_0 = 1\% B_c, 5\% B_c, 15\% B_c$. The legend is the same for both the figures.}
    \label{djrhodirhodiuj}
\end{figure*}

The term $-\left \langle 2 (\partial_i \rho \partial_j \rho) \partial_i u_j \right \rangle_V $ denotes the stretching of density gradient tensor ($\partial_i \rho \partial_j \rho$) by the velocity ($u_j$) along the direction $i$. Plotting the above quantity (see figure \ref{djrhodirhodiuj}(left)) we find that, for a given mixing layer height ($h$), the 3D MRTI (dashed lines) undergoes more stretching than the 2D MRTI (solid lines) for all magnetic field strengths. This tells that the 3D MRTI case has more elongated structures, possibly due to the interchange modes that have HD-like evolution. 

The time scale for the fluid stirring can be calculated by taking the ratio of density gradient norm, $\left \langle (\partial_i \rho)^2 \right \rangle_V $ to $-\left \langle 2 (\partial_i \rho \partial_j \rho) \partial_i u_j \right \rangle_V $. From figure \ref{djrhodirhodiuj}(right), we find that the timescale of stirring is much higher in 2D, i.e., fluid is less stirred in 2D compared to 3D. The quick stirring in 3D is due to the interchange, mixed, and the coupled modes, which evolve faster than the undular modes. 

Comparing across different magnetic field strengths, we find that as the magnetic field strength increase, stirring decreases and stirring time scale increases (for both 2D and 3D). As the magnetic field strength increases, the small-scale turbulent structures diminish suppressing the stirring. Further, the dominance of large scale structures lead to longer stirring time scales.

\subsubsection{Length scales} \label{sec:lengthscales}

\begin{figure*}[t]
    \centering
    \begin{subfigure}[b]{0.4\textwidth}
         \centering
         \includegraphics[width= \linewidth]{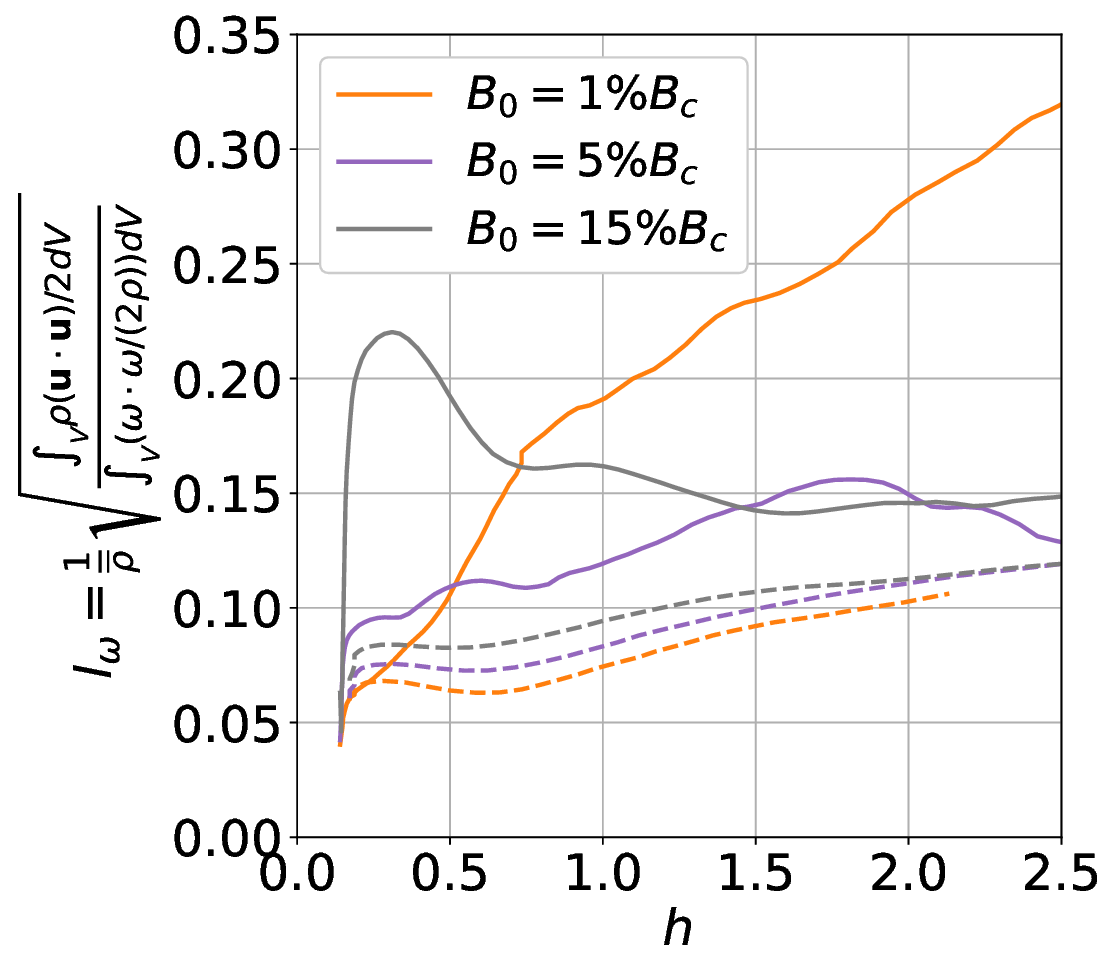}
     \end{subfigure}
    \begin{subfigure}[b]{0.4\textwidth}
         \centering
         \includegraphics[width= \linewidth]{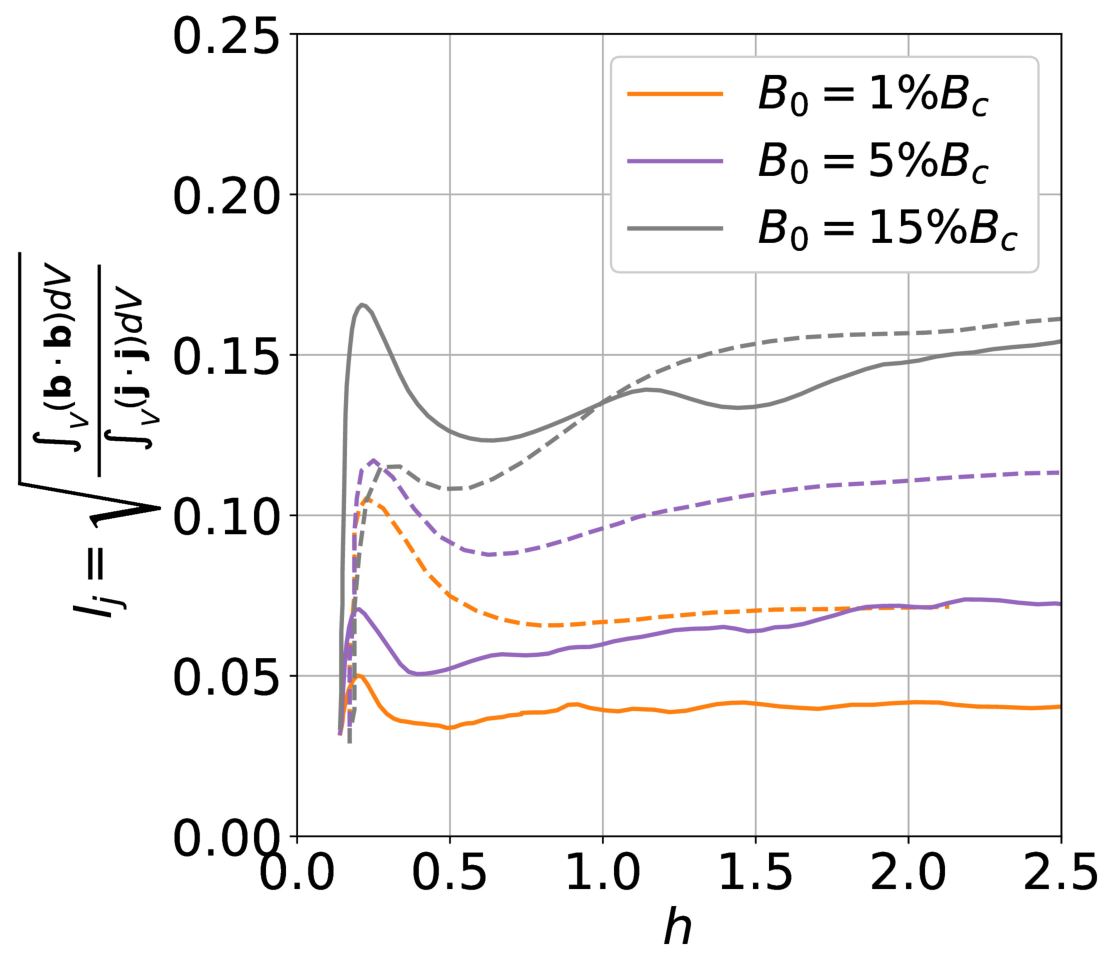}
     \end{subfigure}
    \caption{Variation of: \textit{(left)} vortex length scales $l_\omega$; and \textit{(right)} current length scales $l_\textbf{j}$ (see equation \ref{lomegaeqn}) with mixing layer height for 2D and 3D MRTI of different magnetic field strengths.}
    \label{vorticitylengthscale}
\end{figure*}
From figure \ref{2d_3d}, the structures in the mixing layer appear to be larger in 2D compared to 3D. To determine the characteristic length scales of the vortical structures ($l_\omega$), we find the ratio of TKE to the enstrophy. Similarly, length scales of current is calculated from the ratio of TME to magnetic enstrophy. 
\begin{equation}
    l_\omega = \frac{1}{\Bar{\rho}} \sqrt{\frac{\left \langle \rho \textbf{u} \cdot \textbf{u}/2 \right \rangle_V}{\left \langle \mathbf{\omega} \cdot \mathbf{\omega}/2 \rho \right \rangle_V}}; l_\textbf{j} = \sqrt{\frac{\left \langle \textbf{b} \cdot \textbf{b}/2 \right \rangle_V}{\left \langle \mathbf{j} \cdot \mathbf{j}/2 \right \rangle_V}}.
    \label{lomegaeqn}
\end{equation}
Here $\Bar{\rho} = \frac{\rho_h + \rho_l}{2}$, $\mathbf{\omega} (= \nabla \times \textbf{u})$ is vorticity, $\textbf{j} (= \nabla \times \textbf{b})$ is the current. The TKE and TME are defined as:
\begin{equation}
    \text{TKE} = \left \langle \frac{\rho \textbf{u} \cdot \textbf{u}}{2} \right \rangle_V; \text{TME} = \left \langle \frac{\textbf{b} \cdot \textbf{b}}{2} \right \rangle_V.
    \label{TKE_TME_eqns}
\end{equation}
The enstrophy and magnetic enstrophy are defined as $\left \langle \frac{\omega \cdot \omega}{2 \rho} \right \rangle_V$ and $\left \langle \frac{\textbf{j} \cdot \textbf{j}}{2} \right \rangle_V$, respectively. In figure \ref{vorticitylengthscale} we plot $l_{\omega}$ and $ l_j$ for 2D and 3D MRTI at different magnetic field strengths. The figure exhibits several interesting features. First, in most cases the $l_{\omega}$ is approximately constant. An exception for this is the weak 2D MRTI case, where the $l_{\omega}$ grows linearly with mixing layer height. A potential reason for this could be the vortex merger leading to larger vortical structures with increasing mixing layer. In the intermediate and strong field cases, magnetic field suppresses turbulence for the $l_{\omega}$ to grow like the weak 2D MRTI case. In 3D, on the other hand, vortex stretching prevents the formation of continuously increasing larger vortex structures keeping $l_{\omega}$ approximately constant.

Second, comparing different magnetic field strengths, $l_{\omega}$ increases monotonously with field strength for 3D MRTI. But such a trend is not observed in 2D case. 

Third, $l_j$ is approximately constant with mixing layer height for both 2D and 3D MRTI at all magnetic field strengths. Unlike the $l_{\omega}$, $l_j$ does not increase with mixing layer height in weak 2D MRTI case. 

Fourth, the $l_j$ increases monotonously with magnetic field strength for both 2D and 3D. At low magnetic field strength, the mixing layer is more turbulent allowing magnetic field to reconnect, break, and form small current sheets. As the magnetic field strength increases, the small scales are suppressed by the magnetic field and the mixing layer is increasingly composed of large scale structures with current sheets wrapped around the plumes heads and planks (see figure 9 of \citet{Kallurireco_2025}). 

\vspace{-10pt}
\subsubsection{Section summary}

In summary, the interchange and mixed modes changes MRTI mixing layer significantly. The mixing layer of 2D MRTI with purely undular modes has large-scale vortical structures (possibly due to the inverse cascade phenomenon), high fluid interpenetration and less fluid mixing. The significant interpenetration, accompanied by less mixing, results in sharp jumps in mixing profiles at the boundaries of the mixing layer. On the contrary, the mixing layer of 3D MRTI with undular, interchange and mixed modes has less interpenetration and higher fluid mixing. This leads to a smooth variation of the mixing parameter as we move to the centre of the mixing layer. The 3D case has smaller vortical structures and hence greater stirring with lower stirring time scales due to the presence of interchange modes that are not influenced by magnetic fields and hence have higher turbulence and HD-like growth. The 2D has large vortical length scales, lower stirring with larger stirring time scales.

\subsection{Energy dynamics}
\begin{figure*}[t]
    \centering
    \includegraphics[width = \textwidth]{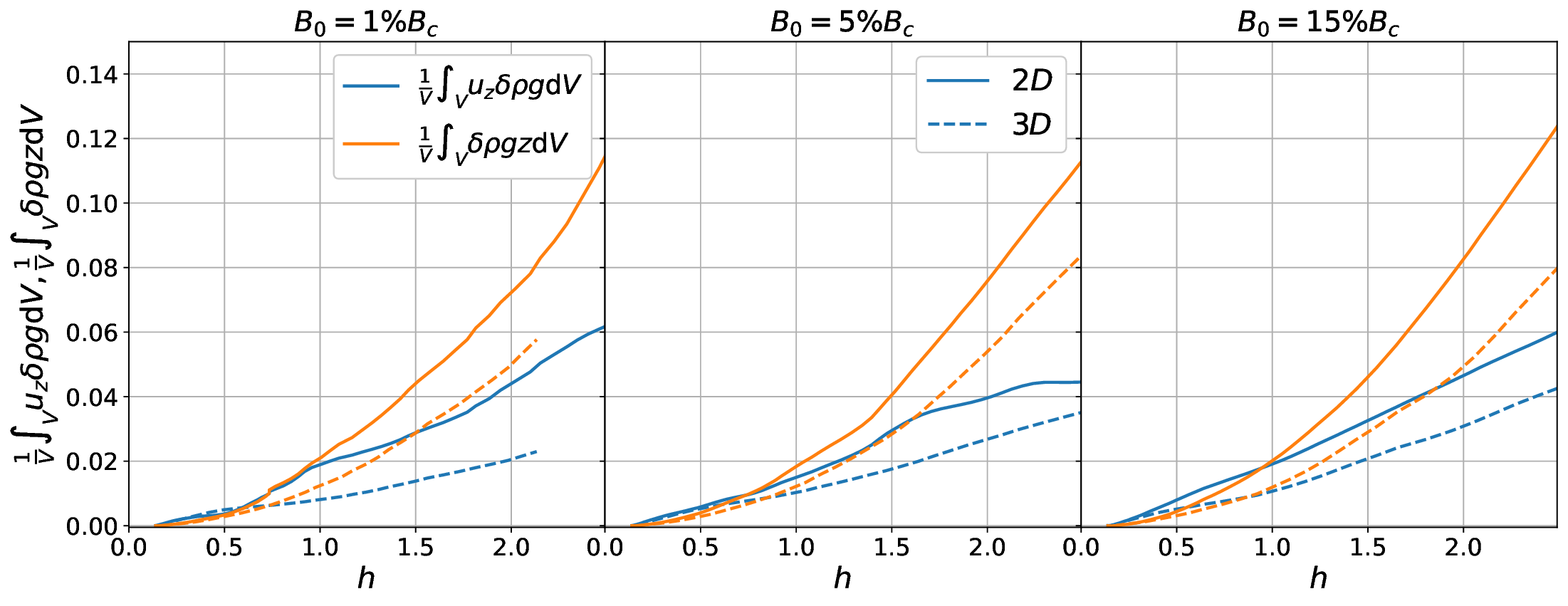}
    \caption{Variation of released gravitational potential energy (GPE), $\left \langle \delta \rho g z \right \rangle_V $; and energy extracted from GPE, $\left \langle u_z \delta \rho g \right \rangle_V $ with mixing layer height (X axis). The solid and dashed lines represent the 2D and 3D cases.}
    \label{GPE}
\end{figure*}

Fundamentally, the evolution and dynamics of mixing layer are related to the energy dynamics. In this section we will understand the differences between the 2D and 3D MRTI from the energy dynamics perspective. 

\begin{figure*}[t]
    \centering
    \includegraphics[width=\linewidth]{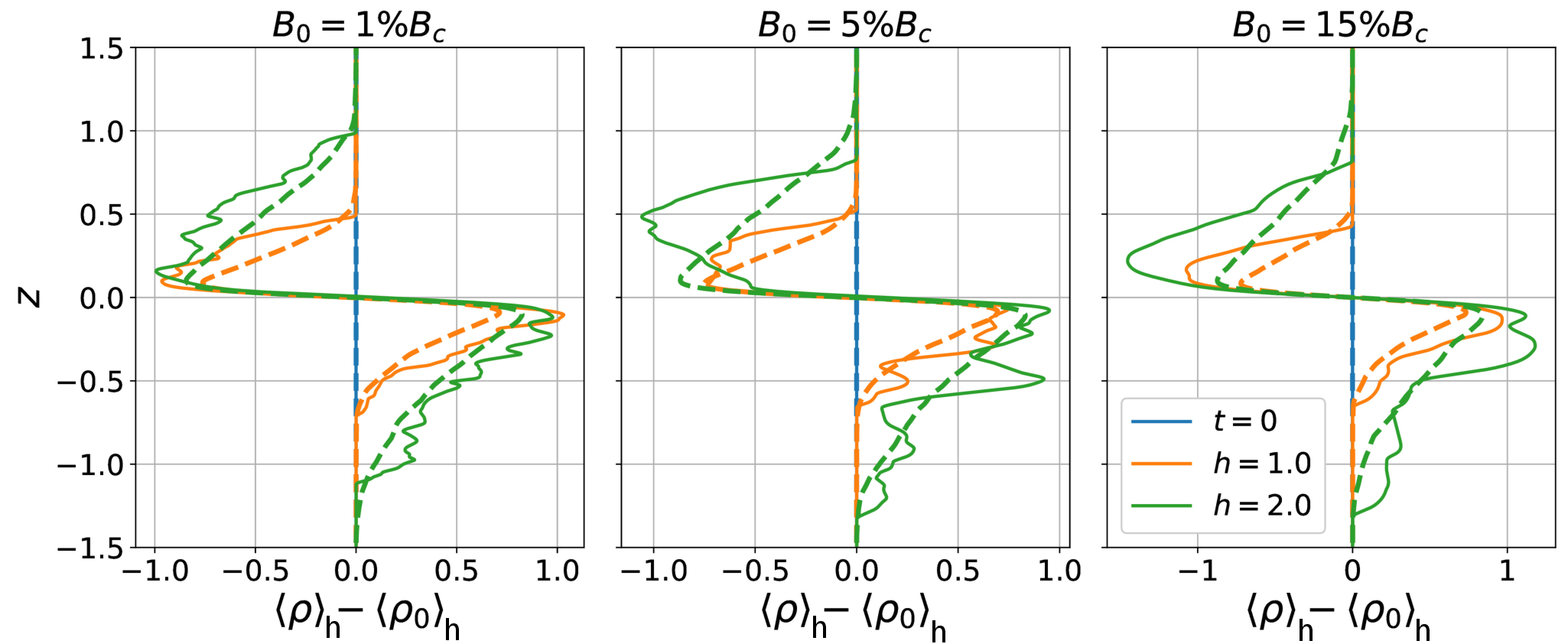}
    \caption{Figure showing the deviation of the homogeneous averaged density perturbation ($\langle \rho - \rho_0 \rangle_h$) along the non-homogeneous direction. The 2D and 3D cases are shown by solid and dashed lines, and for different magnetic field strengths: $B_0 = 1\%$(left); $B_0 = 5\%$(center); $B_0 = 15\%$(right).}
    \label{rhoxy}
\end{figure*}

The energy source in the RTI system is the release of gravitational potential energy (GPE). Plotting the GPE with mixing layer height, see figure \ref{GPE}, we find that the released GPE, calculated as $\left \langle \delta \rho g z \right \rangle_V $, is higher for the 2D MRTI. This is consistent at all magnetic field strengths. To understand this, we plot the homogeneous averaged density perturbation ($\langle \delta \rho \rangle_h \equiv \langle \rho - \rho_0 \rangle_h$) along the non-homogeneous direction ($z$) at fixed mixed layer heights for the 2D and 3D MRTI. The plot, figure \ref{rhoxy}, shows that for a given mixing layer height, the density perturbation is larger in 2D than in 3D. This is due to significant interpenetration and little mixing of the two fluids in 2D (cf. $\S$\ref{sec:Mixing}). This is consistent at all magnetic field strengths, and indicates why released GPE is larger in 2D at all magnetic field strengths.  

\subsubsection{Energy flow in the MRTI system} \label{sec:energyflow}
\begin{figure*}[t]
    \centering
    \begin{subfigure}[b]{\textwidth}
         \centering
         \includegraphics[width= \linewidth]{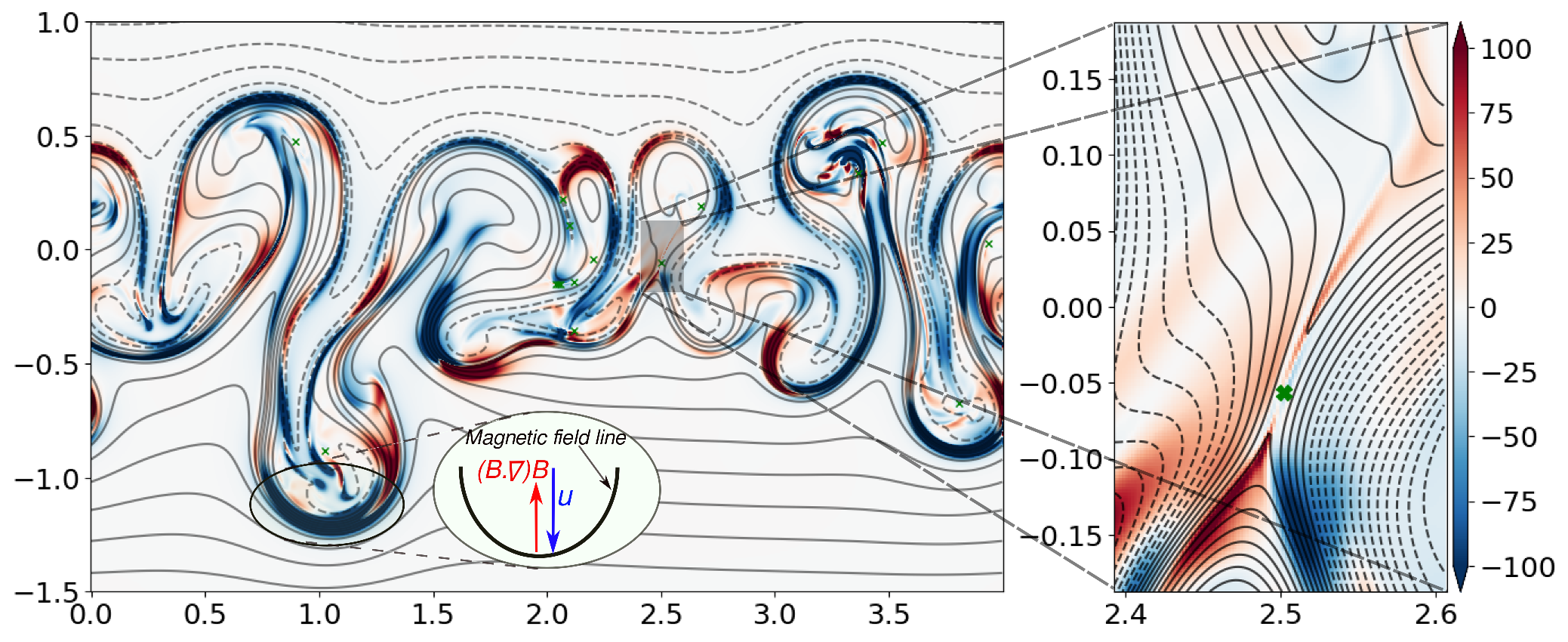}
     \end{subfigure}
    \begin{subfigure}[b]{\textwidth}
         \centering
         \includegraphics[width= \linewidth]{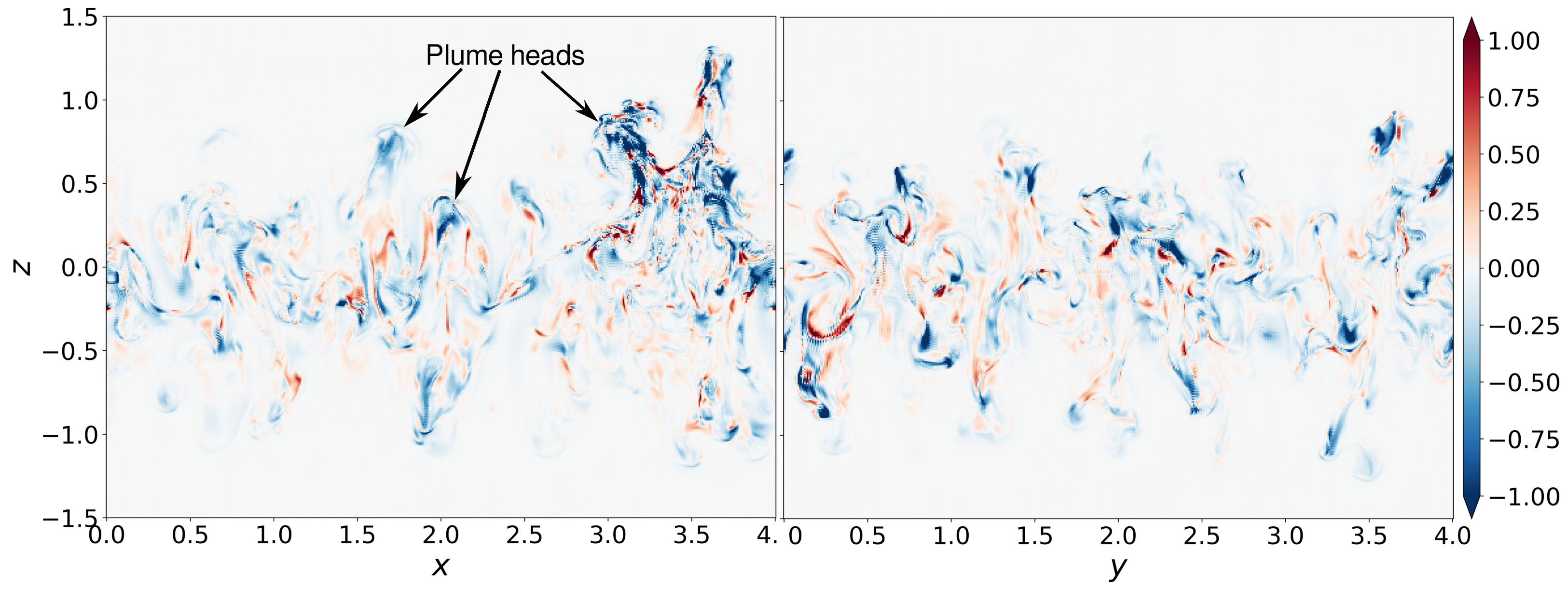}
     \end{subfigure}
    \caption{Instantaneous spatial contour of $\mathbf{u} {\cdot} (\mathbf{B} {\cdot} \nabla) \mathbf{B}$ for 2D (top) and 3D (bottom) MRTI, with magnetic field strength $B_0 {=} 15\%B_c$. In 3D MRTI case, the contours are plotted at mid-$y$ (bottom left) and mid-$x$ (bottom right) planes. \textcolor{ForestGreen}{$\times$} and the black lines in the top panel are the reconnection points and magnetic field lines respectively.}
    \label{ubdb}
\end{figure*}

\begin{figure*}[t]
    \centering
    \includegraphics[width=\linewidth]{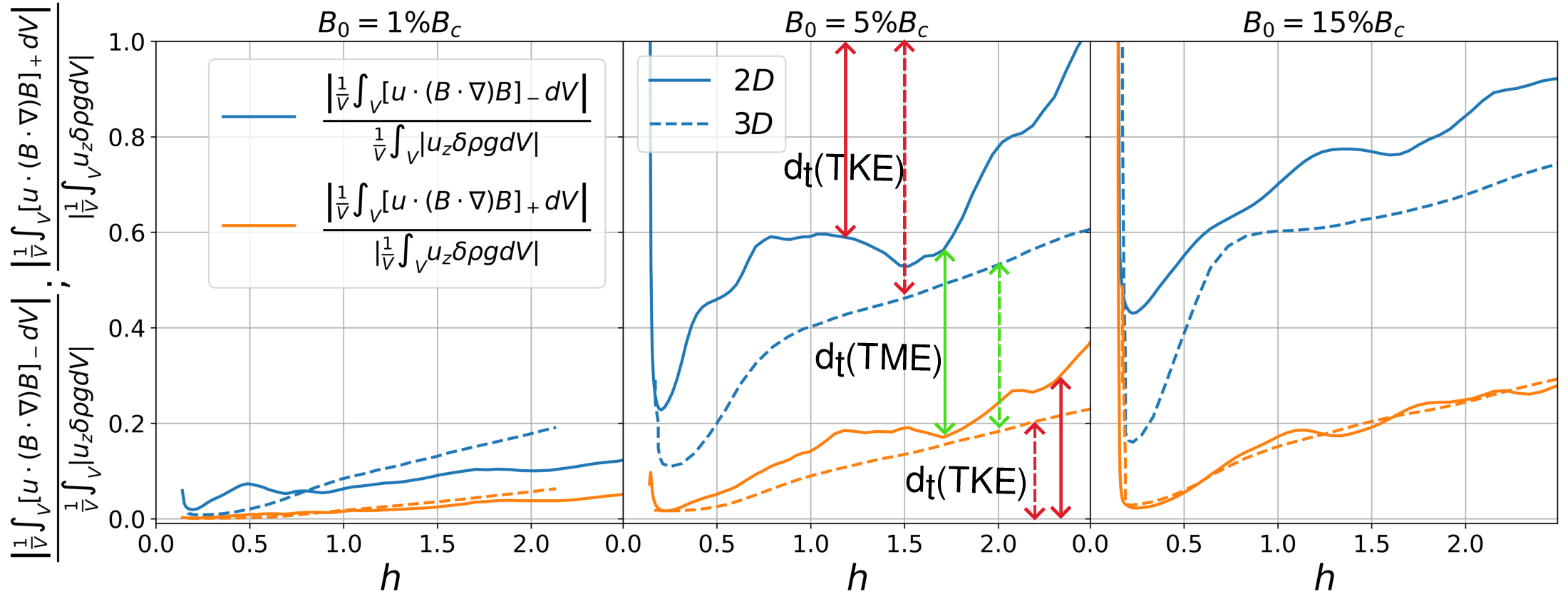}
    \caption{Variation of $\left \langle E_{u \rightarrow b} \right \rangle_V $ and $\left \langle E_{b \rightarrow u} \right \rangle_V $ with mixing layer height $h$. The quantities are normalized by energy extracted from GPE, $\left \langle \delta \rho g u_z \right \rangle_V $. The 2D and 3D MRTI cases are shown by solid and dashed lines.}
    \label{ubdbquantchap3}
\end{figure*}

The flow of energy within the MRTI system can be understood from the TKE equation, given by,
\begin{multline}
    \frac{D}{D t} \left( \rho \frac{\textbf{u} \cdot \textbf{u}}{2} \right) = - (\textbf{u} \cdot \nabla)p + \textbf{u} \cdot \left((\textbf{B} \cdot \nabla) \textbf{B} \right)  \\ - \textbf{u} \cdot (\delta \rho \textbf{g}) + \mathcal{D}_\text{TKE} - \nu \nabla^2 \left( \rho \frac{\textbf{u} \cdot \textbf{u}}{2} \right)
    \label{TKE_eqnchap3}
\end{multline}
The turbulent kinetic energy dissipation term, $\mathcal{D}_\text{TKE}$, will be discussed later in $\S$\ref{sec_energydiss}. The term $- \textbf{u} \cdot (\delta \rho \textbf{g}) (\equiv u_i \delta \rho g \updelta_{i3}$, since $\textbf{g} {=} -g)$ represents the rate of GPE extraction into the kinetic energy through the vertical velocity. A part of this energy is used to deform the magnetic field lines (see figure \ref{ubdb}), which increases magnetic tension and the TME. The larger the magnetic field strength, the greater the magnetic tension and the TME. The converse energy transfer (from magnetic to kinetic energy) also occurs. This energy interplay between the fluid and the magnetic field is given by the term $\mathbf{u} \cdot ((\mathbf{B} \cdot \nabla) \mathbf{B})$. 

Taking volume integral of equation \ref{TKE_eqnchap3} and normalising with released GPE ($-\langle \textbf{u} \cdot (\delta \rho \textbf{g}) \rangle_V$):
\begin{multline}
    \frac{D}{D t} \left( \frac{ \langle \rho \textbf{u} \cdot \textbf{u}/2 \rangle_V}{-\langle \textbf{u} \cdot (\delta \rho \textbf{g}) \rangle_V} \right) = \frac{\langle \textbf{u} \cdot \left((\textbf{B} \cdot \nabla) \textbf{B} \right) \rangle_V}{-\langle \textbf{u} \cdot (\delta \rho \textbf{g}) \rangle_V}  + 1 \\ + \frac{\mathcal{D}_\text{TKE}}{-\langle \textbf{u} \cdot (\delta \rho \textbf{g}) \rangle_V} - \frac{\nu \nabla^2 \left( \rho \textbf{u} \cdot \textbf{u}/2 \right)}{-\langle \textbf{u} \cdot (\delta \rho \textbf{g}) \rangle_V}.
    \label{TKE_eqnchap4}
\end{multline}
Note the use of $\int_V (\textbf{u} \cdot \nabla)p = 0$ for an incompressible system in triply periodic domain. The energy transfer rate from the fluid to the magnetic field is given by the negative $\mathbf{u} \cdot ((\mathbf{B} \cdot \nabla) \mathbf{B})$, and it's corresponding normalized quantity is given by 
\begin{equation}
    E_{u \rightarrow b} = \frac{\left \langle [\mathbf{u} \cdot ((\mathbf{B} \cdot \nabla) \mathbf{B})]_- \right \rangle_V}{\left \langle u_z \delta \rho g \right \rangle_V}.   
\end{equation}
The energy transfer from fluid to magnetic field happens through the deformation of magnetic field lines by the flow, as shown in the inset of figure \ref{ubdb} (top). As the instability evolves, the fluid at the plume heads pushes against the magnetic tension. In this process, the fluid kinetic energy is converted to magnetic energy. Both the 2D and 3D MRTI cases exhibit the energy transfer from the fluid to the magnetic field, and it is evident from the negative $\mathbf{u} \cdot ((\mathbf{B} \cdot \nabla) \mathbf{B})$ in figure \ref{ubdb} at the plume heads. 

The variation of $E_{u \rightarrow b}$ with mixing layer height, blue line in figure \ref{ubdbquantchap3}, shows that the amount of energy expended to deform the magnetic field lines increases with mixing layer height. The proportion of GPE used to deform the magnetic field lines ($E_{u \rightarrow b}$) increases with the magnetic field strength. 

The fraction of the perturbation energy associated with magnetic-field deformation is lower in 3D than in 2D. This can be understood from the additional degree of freedom available to the perturbations in three dimensions. In 2D, the displacement of the fluid is constrained to a plane, so displacing the plasma across the magnetic field necessarily produces bending of the magnetic field lines throughout the perturbed region. In contrast, 3D interchange and mixed modes can partially displace fluid structures in the direction perpendicular to the 2D plane, allowing adjacent magnetic field line bundles to slip past one another rather than requiring all of the field lines to undergo the same degree of bending. For example, consider a magnetic field-line bundle initially located around the plane ($y=L_y/2$). In 3D, a rising structure can be displaced toward ($y=L_y/2 \pm \Delta y$), effectively moving between neighbouring planes. (This is analogous to a balloon rising through a parallel array of strings. Rather than bending every string substantially, the balloon can move between neighbouring strings, locally displacing them and allowing them to slip around the structure.) Consequently, a given fluid displacement can be achieved with less magnetic-field-line bending in 3D than in 2D, resulting in a smaller fraction of the perturbation energy being stored in magnetic-field deformation.

The rate of energy transfer from the magnetic field to the fluid is given by a positive value of $\mathbf{u} \cdot ((\mathbf{B} \cdot \nabla) \mathbf{B})$, and it's corresponding normalized quantity is given by 
\begin{equation}
    E_{b \rightarrow u} = \frac{\left \langle [\mathbf{u} \cdot ((\mathbf{B} \cdot \nabla) \mathbf{B})]_+ \right \rangle_V}{\left \langle u_z \delta \rho g \right \rangle_V}.   
\end{equation}
One such process that releases the energy in the magnetic field to the fluid is magnetic reconnection, see jet like structures emanating from the reconnection point ($\textcolor{ForestGreen}{\times}$) in the figure \ref{ubdb}. A thorough investigation of the role of reconnection highlighting these features in energy conversion was discussed by \citet{Kallurireco_2025}. The energy conversion from magnetic to kinetic energy occurs both in 2D and 3D (regions in red in figure \ref{ubdb}). The contour (figure \ref{ubdb}) shows that the energy conversion from magnetic to fluid (red regions) is much lower than the energy conversion from fluid to magnetic (blue regions). This is evident from figure \ref{ubdbquantchap3} where we also plot the temporal variation of $E_{b \rightarrow u}$ (orange line). 

Similar to $E_{u \rightarrow b}$, the energy converted from magnetic to kinetic energy ($E_{b \rightarrow u}$) increases with mixing layer height, and for a given mixing layer height, the energy converted to kinetic energy is higher in the strong magnetic field case. Surprisingly, $E_{b \rightarrow u}$ is similar in both 3D and 2D. This hints that the energy conversion from magnetic to fluid in 3D simulations could be explained and estimated from 2D simulations.

\subsubsection{Turbulent Kinetic and Magnetic Energies}

\begin{figure*}[t]
    \centering
    \includegraphics[width=\linewidth]{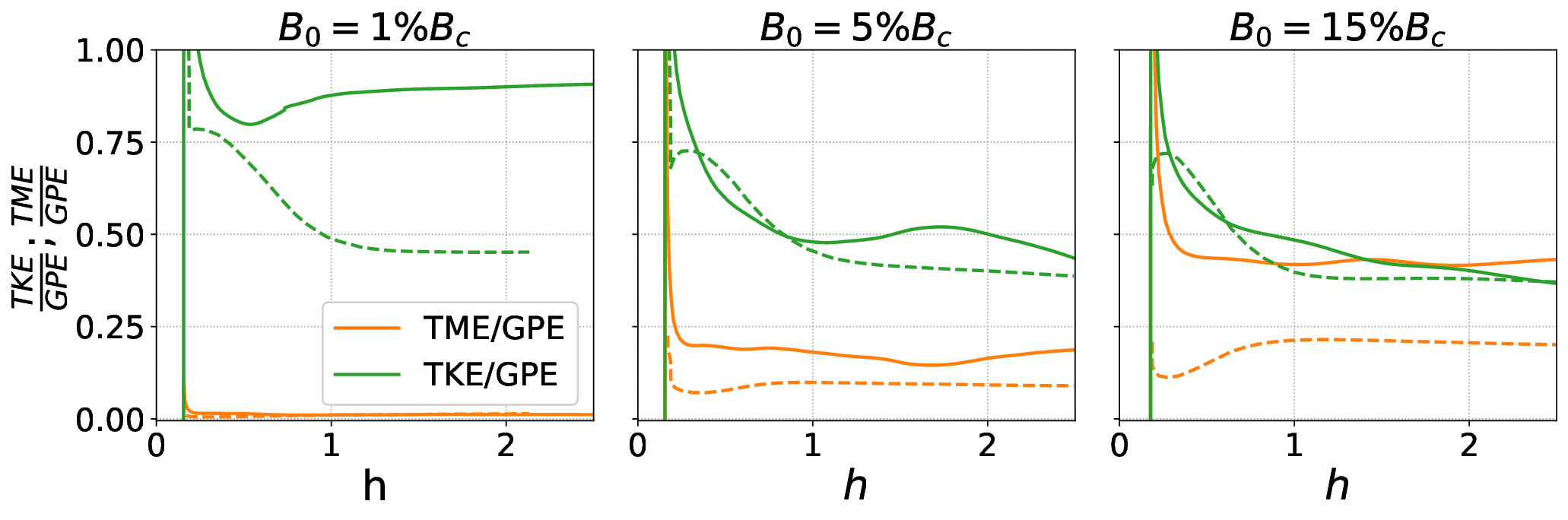}
    \caption{Variation of normalized TKE, TME with mixing layer height for different magnetic field strengths. The 2D and 3D MRTI cases are shown by solid and dashed lines.}
    \label{TKE_TMEchap3}
\end{figure*}
Most natural systems have negligible energy dissipation due to their large Reynolds numbers. In such systems, $E_{u \rightarrow b}$ and $E_{b \rightarrow u}$ can give insight into the TKE and TME in the system. From equation \ref{TKE_eqnchap4}, when $\mathcal{D}_\text{TKE} \approx 0$ and  $\nu \nabla^2 \left( \rho \textbf{u} \cdot \textbf{u}/2 \right) \approx 0$, we get
\begin{equation*}
    \frac{D}{D t} \left( \frac{ \langle \rho \textbf{u} \cdot \textbf{u}/2 \rangle_V}{-\langle \textbf{u} \cdot (\delta \rho \textbf{g}) \rangle_V} \right) =  \frac{\langle \textbf{u} \cdot \left((\textbf{B} \cdot \nabla) \textbf{B} \right) \rangle_V}{-\langle \textbf{u} \cdot (\delta \rho \textbf{g}) \rangle_V}  + 1.
\end{equation*}
\begin{equation}
    \implies \frac{\text{TKE}}{\text{GPE}} =  E_{b \rightarrow u} + (1 - E_{u \rightarrow b}).
\end{equation}
That is the TKE in the system is the sum of $E_{b \rightarrow u}$ and $(1 - E_{u \rightarrow b})$, marked as $d_t (\text{TKE})$ at bottom and top in figure \ref{ubdbquantchap3}(centre). Similarly one can show that the TME depends on the difference between $E_{u \rightarrow b}$ and $E_{b \rightarrow u}$, marked as $d_t (\text{TME})$ in figure \ref{ubdbquantchap3}. Note that, these are the rates of energy injection, hence represented as $d_t (\star)$.

From figure \ref{ubdbquantchap3} we make a few predictions. First, since $d_t(\text{TME})$ increases with increasing magnetic field strength, we expect the TME to increase with field strength. Second, since $E_{u \rightarrow b}$ is similar for 2D and 3D, we expected the TKE to be large and similar for 2D and 3D in the weak magnetic field case. Third, lower energy is expended ($E_{u \rightarrow b}$) to displace the plume in 3D. Consequently, we expect lower TME in 3D.

To confirm these predictions, we plot the TKE and TME (defined in equation \ref{TKE_TME_eqns}) normalised by GPE. As expected, the TME increases with magnetic field strength and the TME is higher in 2D compared to 3D. However, the second prediction does not seem to hold. Figure \ref{TKE_TMEchap3} shows a stark difference between TKE in 2D and 3D. The TKE is $\approx 90\%$ of the GPE in 2D, whereas in 3D it is only $\approx 45\%$ of the GPE. One possible explanation for this could be the assumption of negligible energy dissipation. If the energy dissipation in 3D is significant, a part of the released GPE is lost, reducing the fraction of TKE. We will investigate the energy dissipation in $\S$\ref{sec_energydiss}. 

\vspace{-10pt}
\subsubsection{Energy dissipation} \label{sec_energydiss}
\vspace{-5pt}

\begin{figure*}
    \centering
    \includegraphics[width = \textwidth]{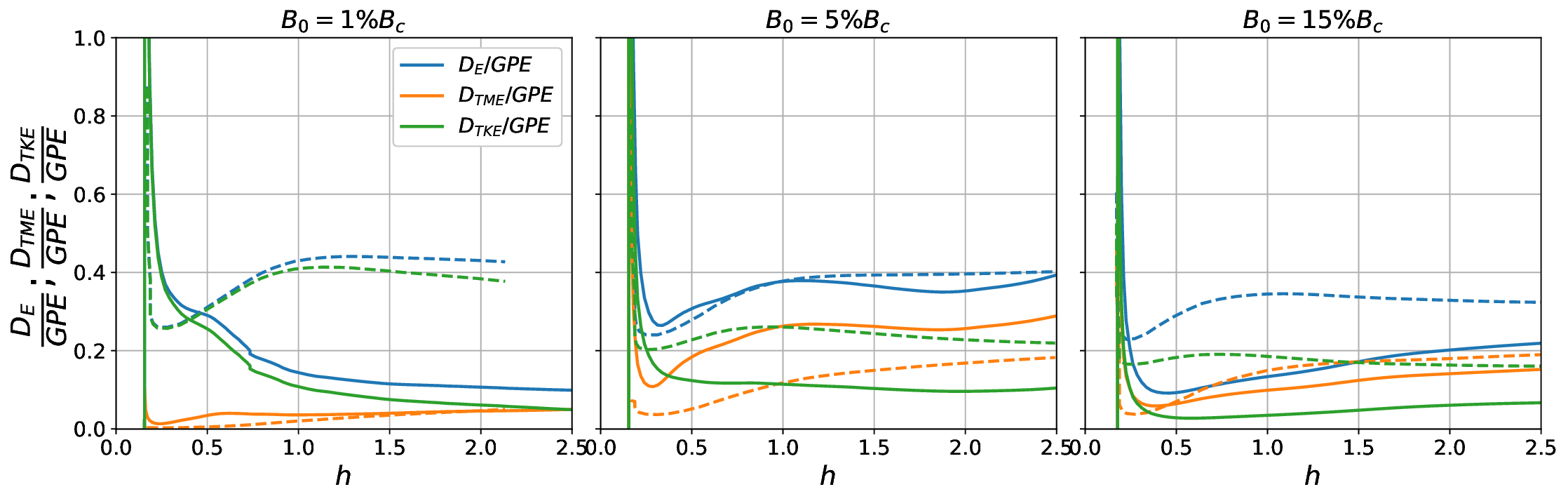}
    \caption{Variation of normalized turbulent kinetic energy dissipation (green line), turbulent magnetic energy dissipation (orange line), and total energy dissipation (blue line) with mixing layer height for the 2D (solid line) and 3D (dashed line) MRTI. All the quantities are normalized by the released GPE.}
    \label{diss}
\end{figure*}

\begin{figure*}
    \centering
    \includegraphics[width = \textwidth]{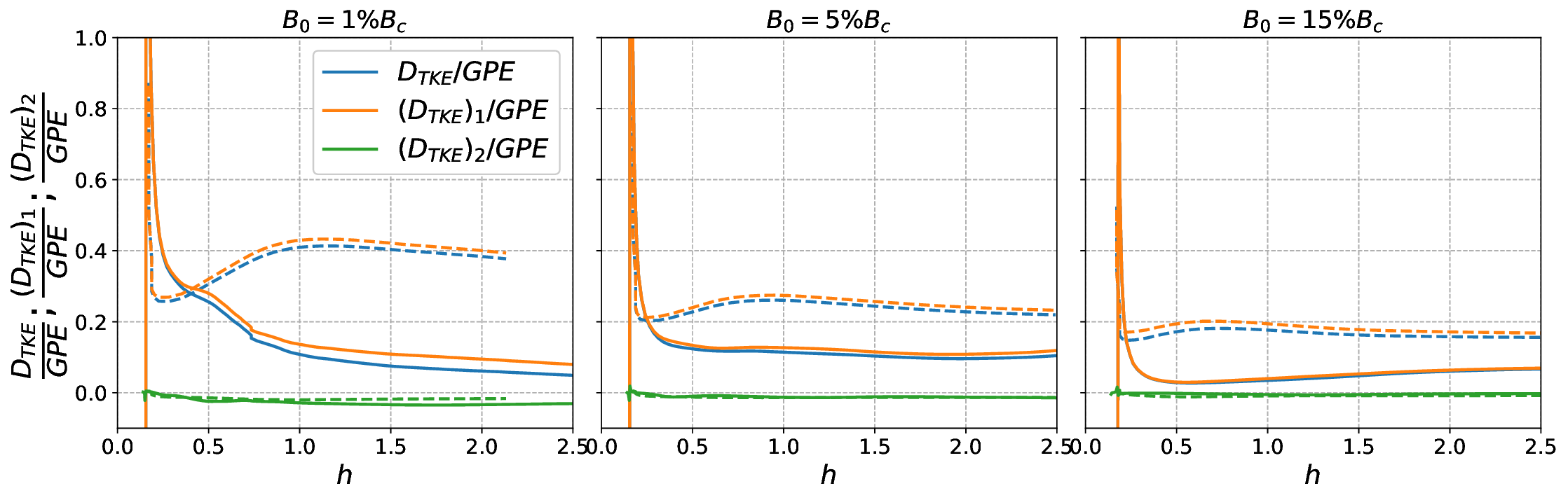}
    \caption{Variation of normalized turbulent kinetic energy dissipation (blue line) and the two terms that make up the dissipation, $(\mathcal{D}_\text{TKE})_1 = \int_0^t \left \langle \nu \rho (\partial_j u_i)^2 \right \rangle_V \mathrm{d}t$ (orange line), $(\mathcal{D}_\text{TKE})_2 = \int_0^t \left \langle (\nu + D) (\partial_j \rho) \left(\partial_j \frac{u_i u_i}{2} \right) \right \rangle_V \mathrm{d}t$ (green line) for the 2D (solid line) and 3D (dashed line) MRTI with mixing layer height. All the quantities are normalized by the released GPE.}
    \label{tke_diss}
\end{figure*}

\begin{figure*}[t]
    \centering
    \includegraphics[width= 0.66\linewidth]{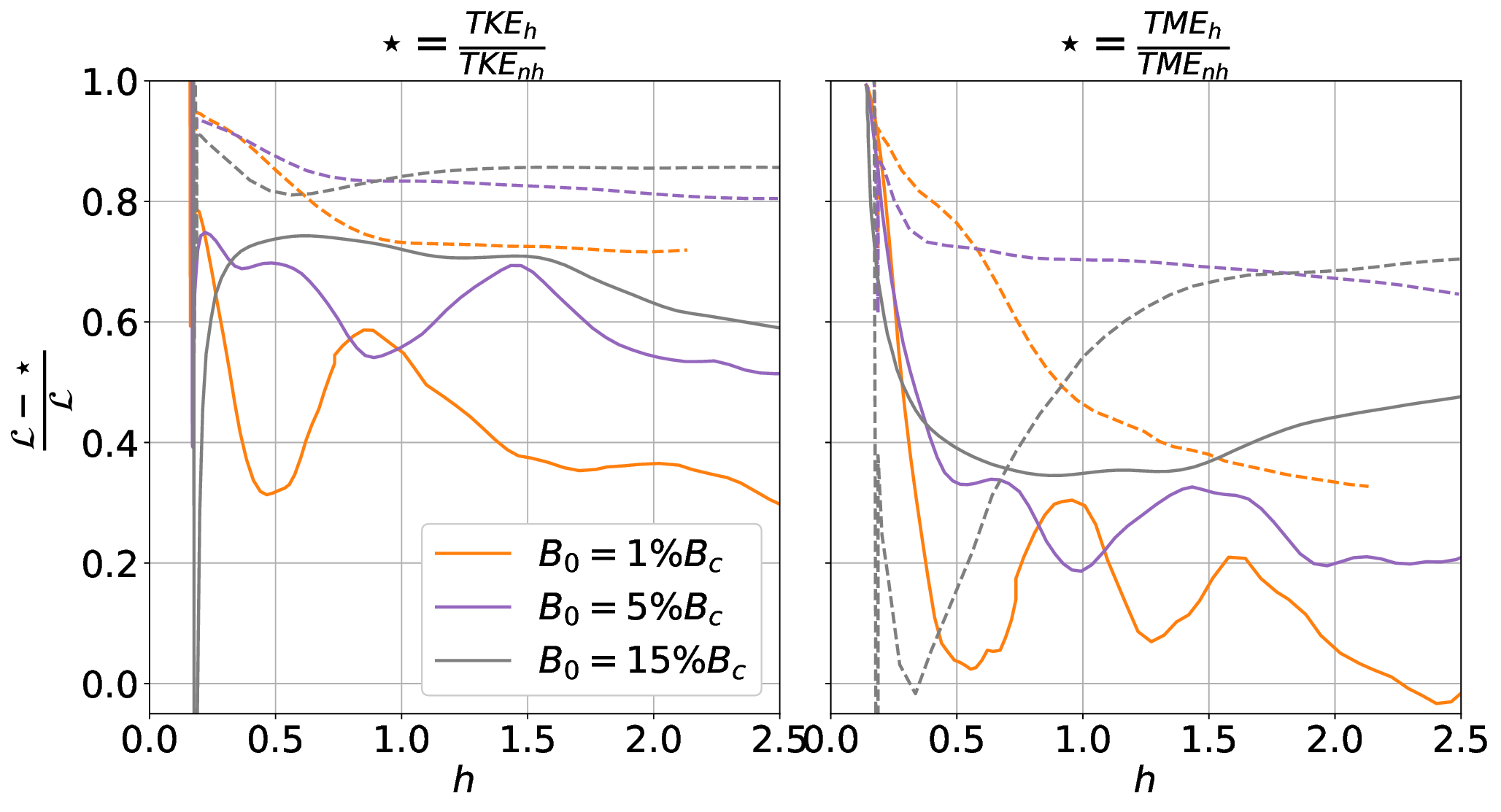}
    \caption{Deviation of (left) TKE and (right) TME from isotropy. 0, 1 are the isotropic and anisotropic limits, respectively. The 2D and 3D MRTI cases are shown by solid and dashed lines.}
    \label{aniso}
\end{figure*}
From energy conservation, the energy balance follows:
{\small
\begin{multline}
    -\underbrace{\left \langle \delta \rho g z \right \rangle_V }_\text{Released GPE} {=} \underbrace{\left \langle \frac{1}{2} \rho u^2 \right \rangle_V}_\text{TKE} {+} \underbrace{\left \langle \frac{1}{2} b^2 \right \rangle_V}_\text{TME} {+} \underbrace{{\int_0^t} {\left \langle \eta (\partial_j b_i)^2 \right \rangle_V \mathrm{d}t}}_\text{TME dissipation $(\mathcal{D}_\text{TME})$} \\ {+} \underbrace{{\int_0^t} {\left \langle \left[ \nu \rho (\partial_j u_i)^2 {+} (\nu {+} D) (\partial_j \rho) \left(\partial_j \frac{u_i u_i}{2} \right) \right] \right \rangle_V \mathrm{d}t}}_\text{TKE dissipation $(\mathcal{D}_\text{TKE})$}.
    \label{energybalance}
\end{multline}}
Hence, when the TKE is significantly lower than what is expected, we believe it could be due to the significant energy dissipation. Calculating the energy dissipation using the formula in equation \ref{energybalance}, we find that 3D MRTI with weak magnetic field has significant TKE dissipation (see figure \ref{diss}), as predicted. 

Figure \ref{diss} shows the total normalized energy dissipation for different magnetic field strengths. In the weak-field case ($B_0 {=} 1\%B_c$), dissipation in 2D (solid blue line) is only ${\approx}10\%$ of GPE, but nearly ${\approx} 35\%$ in 3D (dashed blue line). Similarly, the energy dissipation is also high for other field strengths. We further delve into the two terms of TKE dissipation: $(\mathcal{D}_\text{TKE})_1 = \int_0^t \left \langle \nu \rho (\partial_j u_i)^2 \right \rangle_V \mathrm{d}t$, which arises from velocity gradients (and thus vorticity), and $(\mathcal{D}_\text{TKE})_2 = \int_0^t \left \langle (\nu + D) (\partial_j \rho) \left(\partial_j \frac{u_i u_i}{2} \right) \right \rangle_V \mathrm{d}t$, which is due to the density gradients. From Figure \ref{tke_diss}, we find that $(\mathcal{D}_\text{TKE})_2$ has negligible contribution irrespective of the magnetic field strengths in both 2D and 3D. Thus, the primary sources of energy dissipation in both 2D and 3D are $(\mathcal{D}_\text{TKE})_1$ and $\mathcal{D}_\text{TME}$, indicating that dissipation is dominated by gradients in velocity and magnetic field, that is, by vorticity and current sheets. 

The discrepancies observed between the predictions from figure \ref{ubdbquantchap3} and observations from figure \ref{TKE_eqnchap3} can also be explained from the characteristic vorticity and current length scales ($l_\omega, l_j$). When $l_\star, \star = \{\omega, j\}$ scales are large, energy resides in large-scale structures that dissipate slowly, resulting in lower total dissipation. Conversely, small $l_\star$ imply energy is concentrated at finer scales, leading to stronger dissipation. The characteristic vorticity length scale, $l_\omega$, is as defined in equation \ref{lomegaeqn}. From figure \ref{vorticitylengthscale}, we find that $l_\omega$ is smaller in 3D than in 2D, especially in the weak-field regime (for $B_0 = 1\%B_c$ at $h = 2.0$, $\frac{(l_\omega)_{2D}}{(l_\omega)_{3D}} {\approx} 2.75$). This explains the larger TKE dissipation observed in 3D, and as expected $\frac{(\mathcal{D}_\text{TKE})_{3D}}{(\mathcal{D}_\text{TKE})_{2D}} {\equiv} \left( \frac{(l_\omega)_{2D}}{(l_\omega)_{3D}} \right)^2 {\approx} 8$. The 3D MRTI has more small-scale vorticity structures, which dissipate more efficiently. In contrast, the weak-field case of 2D MRTI has large $l_\omega$, i.e., the mixing layer is dominated by large-scale vortices and consequently has low dissipation. This is visually corroborated in Figure \ref{2d_3d}, where 2D flows show coherent large vortices, while 3D flows are characterized by more diffuse, smaller-scale structures that enhance energy dissipation.

The characteristic length scales of current sheets, denoted by $l_j$, can be estimated based on equation \ref{lomegaeqn}. Figure \ref{vorticitylengthscale}(right) shows the values of $l_j$ for 2D and 3D MRTI cases across different magnetic field strengths. We find that $l_j$ is marginally larger in 3D than in 2D, which corresponds to a slightly lower TME dissipation in 3D. While $l_j$ is generally smaller than the vorticity length scale $l_\omega$, the overall contribution of TME dissipation ($\mathcal{D}_\text{TME}$) remains subdominant compared to TKE dissipation ($\mathcal{D}_\text{TKE}$), primarily due to the lower magnitude of TME in the system. Indeed $l_\omega$ and $l_j$ are effected by the type of wave modes, which was already discussed in $\S$\ref{sec:lengthscales}.

It is also possible that the difference in the energy dissipation between 2D and 3D is due to the difference in the grid resolution; a detailed discussion on the consequences of this difference is presented in $\S$\ref{sec:grid}. 

\subsubsection{Energy anisotropy}

Another important aspect of the energy dynamics is distribution of energy among different components. We quantify anisotropy by calculating 
\begin{equation*}
    \frac{\mathcal{L} - \star}{\mathcal{L}} \text{, where} \star {=} \frac{\text{TKE}_h}{\text{TKE}_{nh}} \left( \frac{\text{TME}_h}{\text{TME}_{nh}} \right).
\end{equation*}
$\mathcal{L}$ is set to 2 for 3D and 1 for 2D. Note that, 
\begin{equation*}
    \text{TKE}_{nh} {=} \frac{1}{2} \rho u_z^2 \text{ and TKE}_h = \left( \frac{1}{2} \rho u_x^2 + \frac{1}{2} \rho u_y^2 \right).    
\end{equation*}
Similarly, 
\begin{equation*}
    \text{TME}_{nh} {=} \frac{1}{2} b_z^2 \text{ and TME}_h = \left( \frac{1}{2} b_x^2 + \frac{1}{2} b_y^2 \right).
\end{equation*}
When the energy is perfectly distributed across all components of TKE or TME, $(\mathcal{L} - \star)/\mathcal{L} = 0$. In the limit of no TKE or TME along the homogeneous directions $(\mathcal{L} - \star)/\mathcal{L} \rightarrow 1$.

From figure \ref{aniso}, we find that the energy redistribution across different components is closer to the isotropic limit for TME than TKE. A possible reason for the relatively well-distribution of TME is due to the extraction of energy from $u_i$ into all components of the Maxwell stress tensor $(b_i b_j)$, cf. rate of energy extraction term, $u_i \partial_j (b_j b_i)$. On the contrary, the TKE energy is extracted from GPE only along the non-homogeneous direction, cf. the rate of energy extraction term, $u_i \delta \rho g \updelta_{i3}$. Comparing the 2D and 3D cases, the 2D MRTI is more isotropic than the 3D. For both 2D and 3D, TKE and TME, the anisotropy increases with magnetic field strength.

\subsubsection{Section summary}
Summarizing the section, we found that the GPE released is much higher in 2D, compared to 3D. This is due to the significant dispersion of fluid with minimal mixing in 2D, as opposed to the 3D case, where the fluid is more mixed and less dispersed. A part of the released GPE is used to deform the magnetic field lines, thus increasing the magnetic tension and TME in the system. We find that the proportion of GPE energy used to deform magnetic field lines is lower in 3D, compared to 2D. This is due to the presence of mixed modes, that does not necessarily bend the field lines. However, the proportion of energy converting from TME to TKE is approximately same in 2D and 3D. The mixing layer in 3D is characterized by small-scale vortical structures that dissipate significantly, but the 2D MRTI mixing layer has large-scale vortical structures and hence lower dissipation. Owing to the differences in the energy dissipation, the TKE is higher in 2D compared to 3D. We also find that the TKE is highly anisotropic, particularly in 3D, with significant energy in the nonhomogeneous direction. Relatively, the TME is less anisotropic since the energy from fluid is distributed among the all components of Maxwell stress tensor, unlike TKE where energy is deposited into the non-homogeneous component to be then redistributed to others.

\subsection{Self-similarity of non-linear MRTI}
\begin{figure*}[t]
    \centering
    \includegraphics[width=0.33\linewidth]{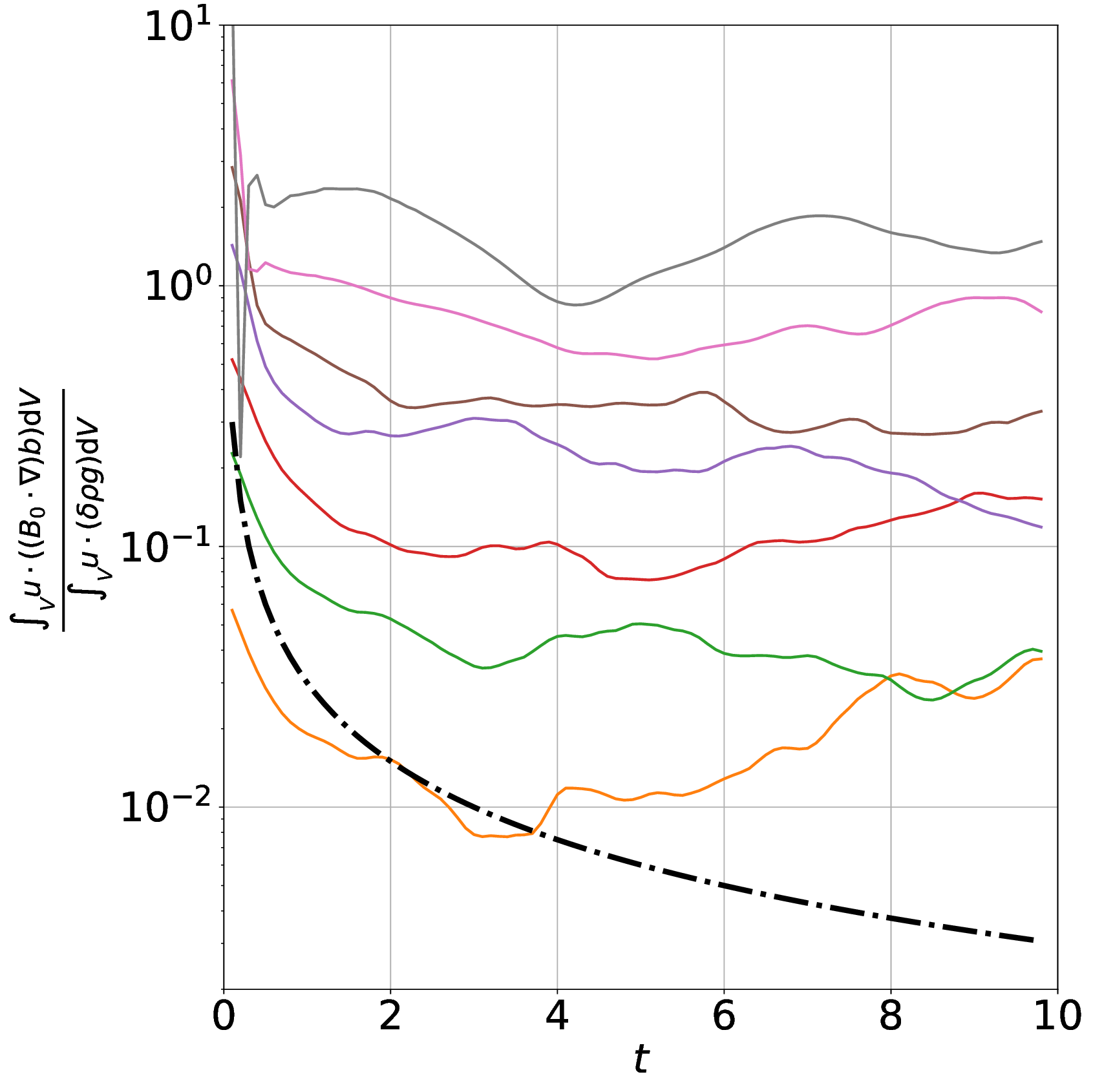}
    \includegraphics[width=0.33\linewidth]{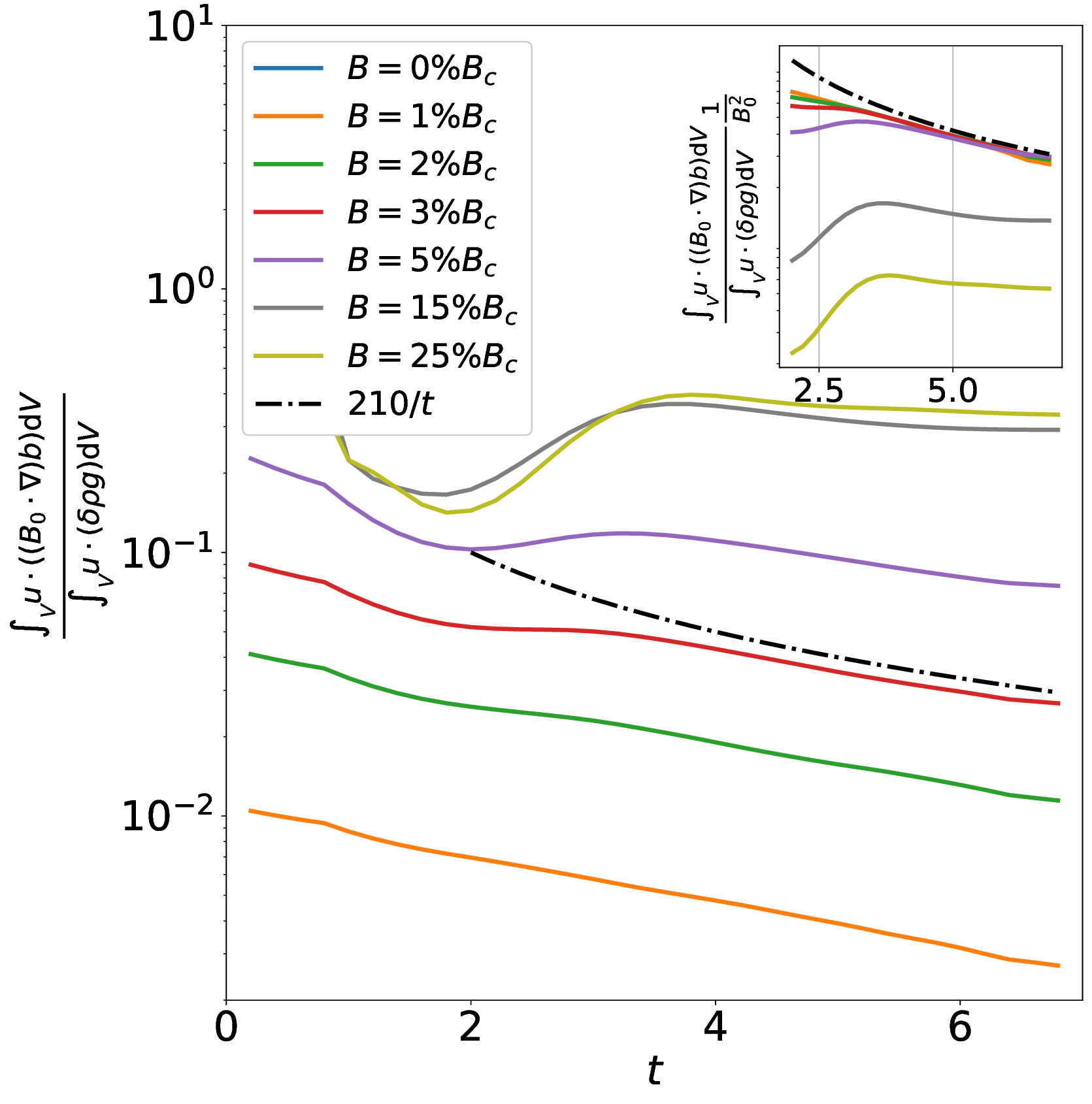}
    \caption{Temporal variation of $\left \langle u \cdot ((B_0 \cdot \nabla) b) \right \rangle_V/ \left \langle u \cdot (\delta \rho g) \right \rangle_V$ for 2D \textit{(left)} and 3D \textit{(right)} MRTI. The legend is same for both the left and right figures. The legend for the left figure is same as figure \ref{hAgh}(left).}
    \label{B0udb_rhov}
\end{figure*}

\begin{figure*}[t]
    \centering
    \includegraphics[width=0.33\linewidth]{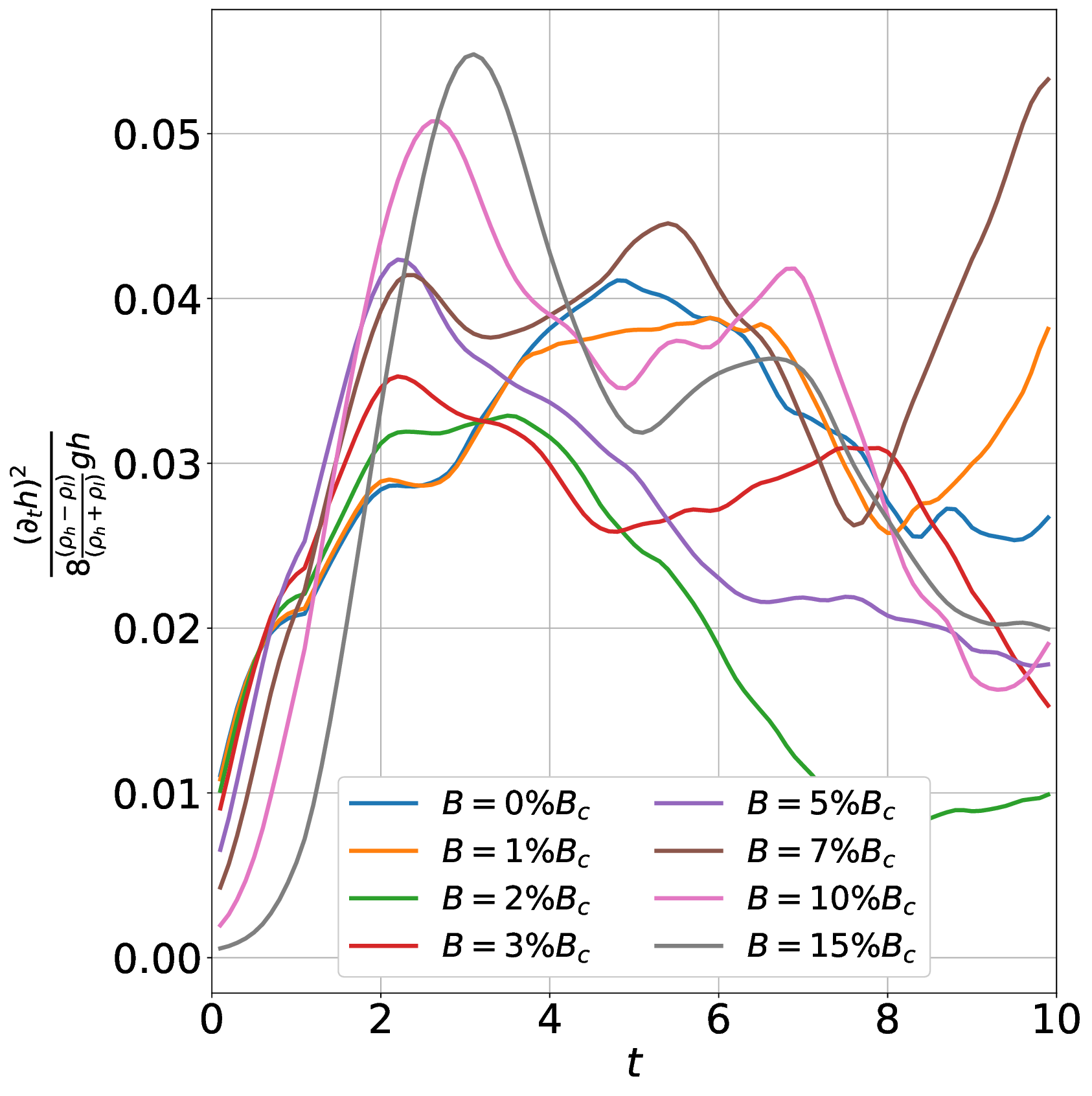}
    \includegraphics[width=0.33\linewidth]{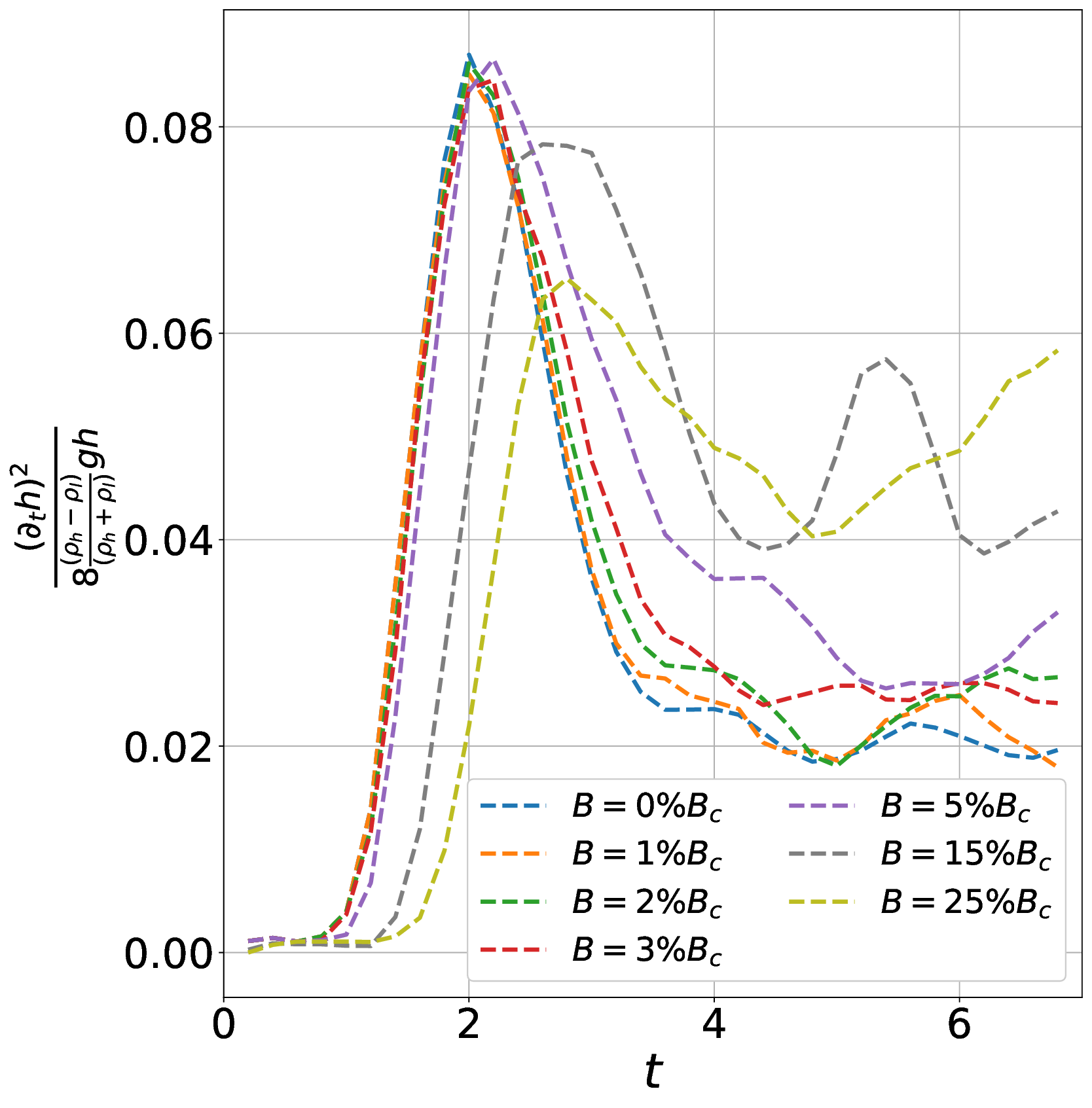}
    \caption{Temporal variation of $\frac{(\partial_t h)^2}{8 \frac{\rho_h - \rho_l}{\rho_h + \rho_l} gh }$ for 2D \textit{(left)} and 3D \textit{(right)} MRTI}
    \label{hAgh}
\end{figure*}

One of the most important consequences of the limited wave modes is reflected in the nature of instability evolution. The broad question of interest here is, does MRTI mixing layer evolve in a self similar fashion in 2D and 3D? The self similar growth of mixing layer is a property where the profile of homogeneous averaged mixing layer at an arbitrary time is the same as the solution at any other time when scaled appropriately, i.e., the height of the mixing layer is expanding in proportion to time. In the case where the magnetic field is aligned parallel to the interface, the 2D MRTI case only has undular modes. The field alignment necessitates the magnetic field lines to be bent (and overcome magnetic tension) through out the instability evolution. Thus the magnetic field has significant influence in the 2D case at all time. However, in the 3D counterpart, the plumes can rise displacing the magnetic field lines (as explained in $\S$\ref{sec:energyflow}). Further the presence of interchange, mixed, and coupled modes, dominate the instability evolution diminishing the influence of imposed magnetic field. To evidence this, we consider the imposed magnetic field term and the gravity term. Indeed from figure \ref{B0udb_rhov}, we find that the ratio of these two quantities decreases in 3D, but remains constant in 2D, backing the arguments that imposed magnetic field has diminishing influence relative to other non-linear terms in 3D, but always has significant influence in 2D. By multiplying the $y$-axis with $1/B_0^2$, in the inset of figure \ref{B0udb_rhov}(right) we show that the $1/t$ variation is valid for all magnetic field strengths.

The recent study by \citet{kalluri2025_self-similarity} proved that the MRTI system approaches towards HD self similarity as the imposed magnetic field term is dominated by the other non-linear terms. However, plotting the ratio of imposed magnetic field term and the GPE term, see figure \ref{B0udb_rhov}, we find that the influence of imposed magnetic field term and GPE term decreases in 3D, but remains approximately constant in 2D MRTI case. That is, our simulations suggest that the 2D MRTI system may achieve self similarity of a different form that is not HD self-similarity. To further clarify that HD self-similarity is not achieved, we calculate $(\partial_t h)^2/\left(8 \frac{\rho_h - \rho_l}{\rho_h + \rho_l} gh \right)$ which should be approximately constant for a HD self-similar system. From figure \ref{hAgh}, we find that the above quantity is approximately constant in 3D but has significant variation in 2D. This strengthens the argument that the 2D MRTI system may not exhibit HD self-similarity like the 3D case. It is important to note that the above does not altogether remove the possibility of self-similarity in 2D, because magnetohydrodynamic self-similarity of a different form is still possible. 

\section{Discussion} \label{sec:discussion}

\subsection{Grid resolution} \label{sec:grid}
\begin{figure*}
    \centering
    \includegraphics[width=\linewidth]{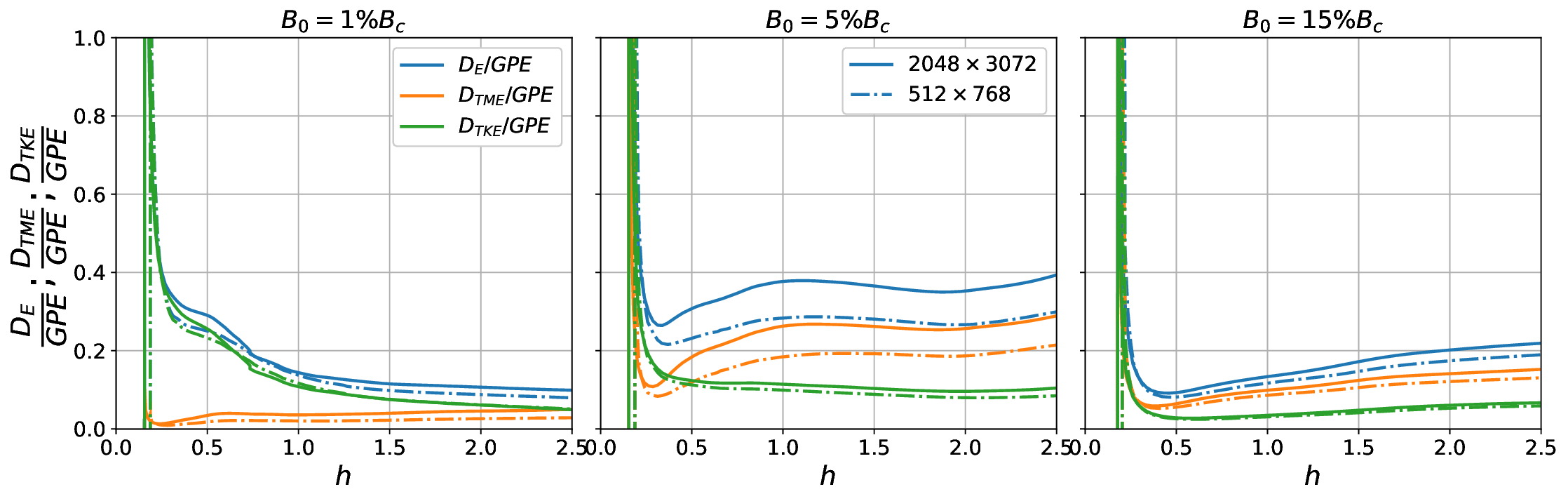}
    \caption{Comparison of normalized turbulent kinetic energy dissipation (green line), turbulent magnetic energy dissipation (orange line), and total energy dissipation (blue line) for different resolutions of 2D MRTI: solid line shows the $2048 \times 3072$ case and dash-dot line shows the $512 \times 768$ case.}
    \label{disstest}
\end{figure*}

\begin{figure*}
    \centering
    \includegraphics[width=\linewidth]{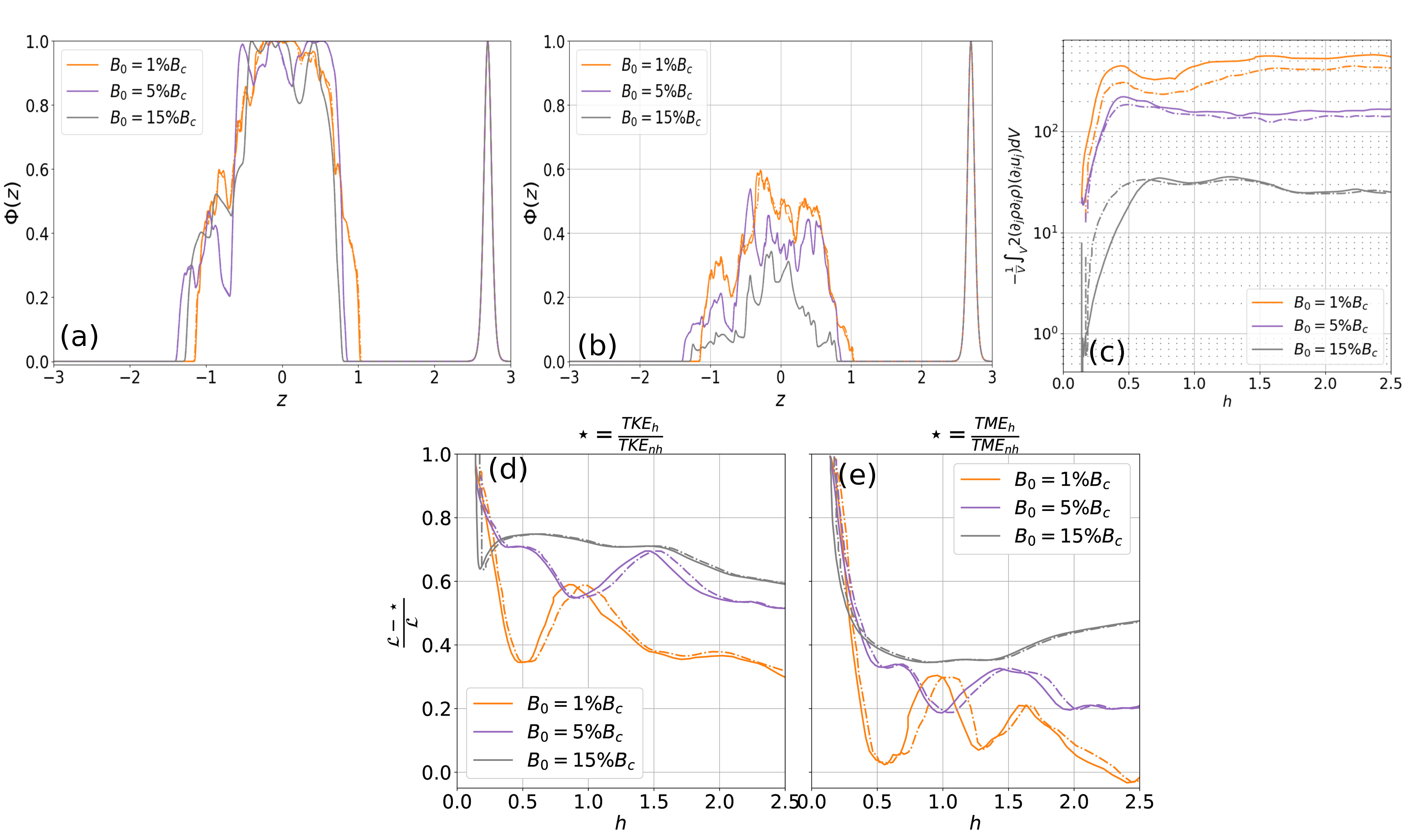}
    \caption{Comparison of: (a) interpenetration, (b) mixing, (c) stirring, (d, e) TKE and TME anisotropy for different resolutions of 2D MRTI: solid line shows the $2048 \times 3072$ case and dash-dot line shows the $512 \times 768$ case.}
    \label{resolcomp}
\end{figure*}

The study explored how the results of 2D and 3D MRTI are different, starting from mixing layer characteristics to energy dynamics like released GPE, TKE, TME, and energy dissipation. A key aspect of the current study is the use of different grid resolutions for the 2D and 3D MRTI. The 2D case uses $2048 \times 3072$, while the 3D case has $512 \times 512 \times 768$ grid points. This is motivated from using as high resolution as possible so that the small scales are well resolved, while balancing the computational constraints. Nevertheless, owing to difference in resolution, there is a need to evidence that the conclusions are resolution independent. Hence we run 2D MRTI simulations at resolution $512 \times 768$ to show that the results are approximately the same. The quantities energy dissipation, energy anisotropy, interpenetration, mixing and stirring for the $2048 \times 3072$ case (solid line) and $512 \times 768$ case (dash-dot line), are plotted in figures \ref{disstest} and \ref{resolcomp}, respectively. The plots show that the two cases have similar variation with time, quantitatively and qualitatively, thus confirming that the differences between 2D and 3D are predominantly due to the differences in dimensionality.

\begin{figure*}[t]
    \centering
    \includegraphics[width=0.66\linewidth]{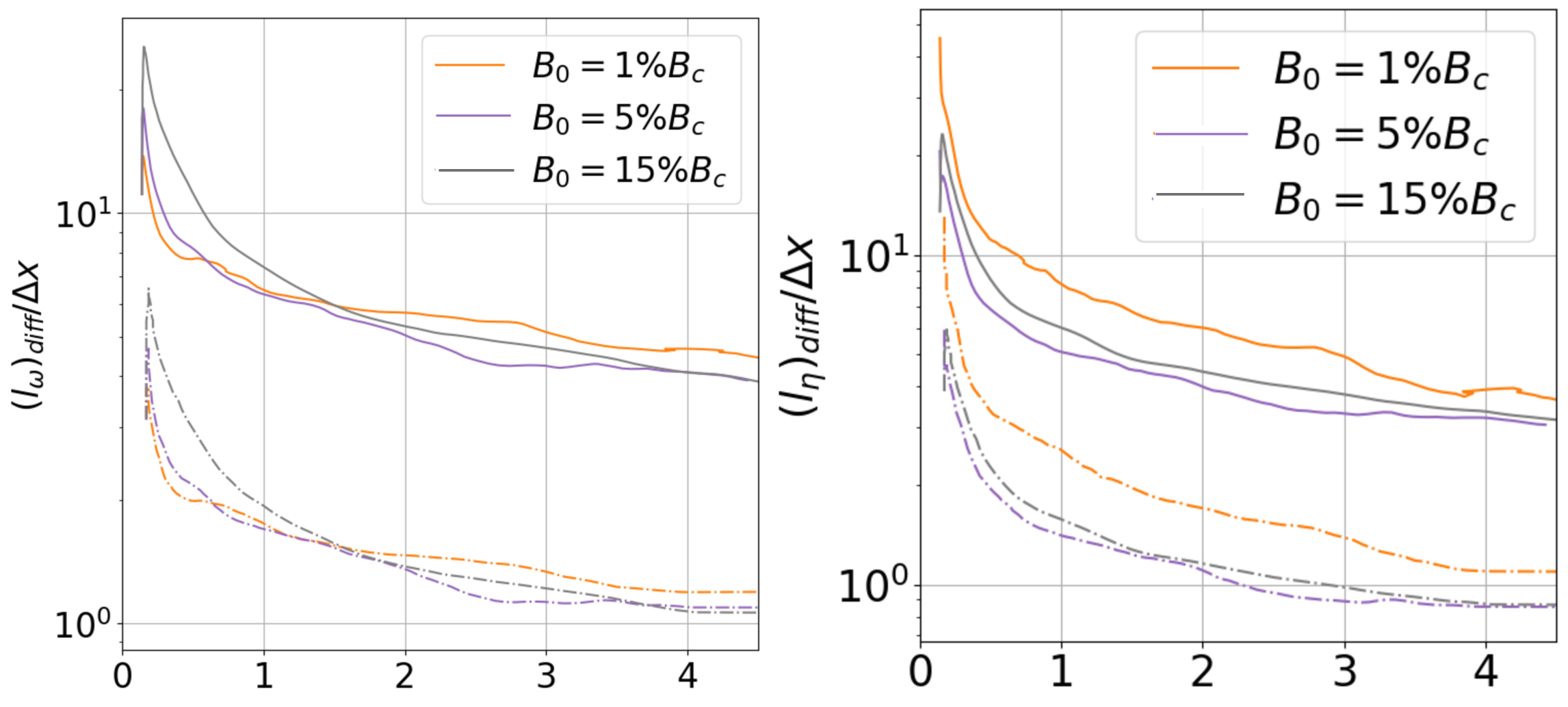}
    \caption{Variation of normalized vorticity and current length scales calculated based on the diffusive scales (normalized with grid scale) with mixing layer height for 2D (solid lines) and 3D (dashed lines) MRTI, for different magnetic field strengths, $B_0 = 1\% B_c, 5\% B_c, 15\% B_c$.}
    \label{l}
\end{figure*}

In the weak field case, where the magnetic field has little dynamical influence, MRTI behaves similarly to HD RTI. Our weak-field results ($B_0 = 1\% B_c$) indeed show good agreement with earlier HD RTI studies \citep{cabot2006comparison, YOUNG_TUFO_DUBEY_ROSNER_2001}. For example, the 2D HD RTI case has higher GPE and TKE compared to the 3D counterpart. The HD RTI studies also reported minimal energy dissipation in 2D and significantly higher dissipation in 3D, as observed in our weak-field MRTI results. This further supports that the results are negligibly affected by the differences in 2D and 3D resolution. Note that the current and vorticity length scales based on global RMS values are well resolved in all our simulations. Further, the smallest current and vorticity length scales are estimated based on diffusion parameters using $(l_\omega)_{diff} = \left( \frac{\nu^3}{\int_t \left \langle \nu \rho (\partial_j u_i)^2 \right \rangle_V \mathrm{d}t} \right)^{1/4}$, $(l_\eta)_{diff} = \left( \frac{\eta^3}{\int_t \left \langle \eta (\partial_j b_i)^2 \right \rangle_V \mathrm{d}t} \right)^{1/4}$. For most part of the instability evolution, $(l_\omega)_{diff}$ and $(l_j)_{diff}$ are well resolved by at least two grid lengths ($2 \Delta x$) in 2D and by atleast one grid lengths ($\Delta x$) in our 3D simulations, as shown in figure \ref{l}. Towards the end, however, they become marginally resolved.

\subsection{Initial perturbations} \label{sec:perturbcomp}
The current 3D simulations start with undular, mixed and interchange modes, whereas the 2D case only supports undular modes. We are thus able to explore the role that
interchange and mixed modes play in MRTI flows. If the 3D simulations were to be perturbed solely with undular modes, MRTI would be expected to evolve to be qualitatively and quantitatively similar to the current 2D simulations. Alternately, if the perturbation contains purely interchange modes, it would be expected to evolve like a purely HD-RTI case, since interchange modes remain entirely unaffected by the magnetic field.

\begin{figure*}[t]
    \centering
    \includegraphics[width=0.32\linewidth,height=5cm]{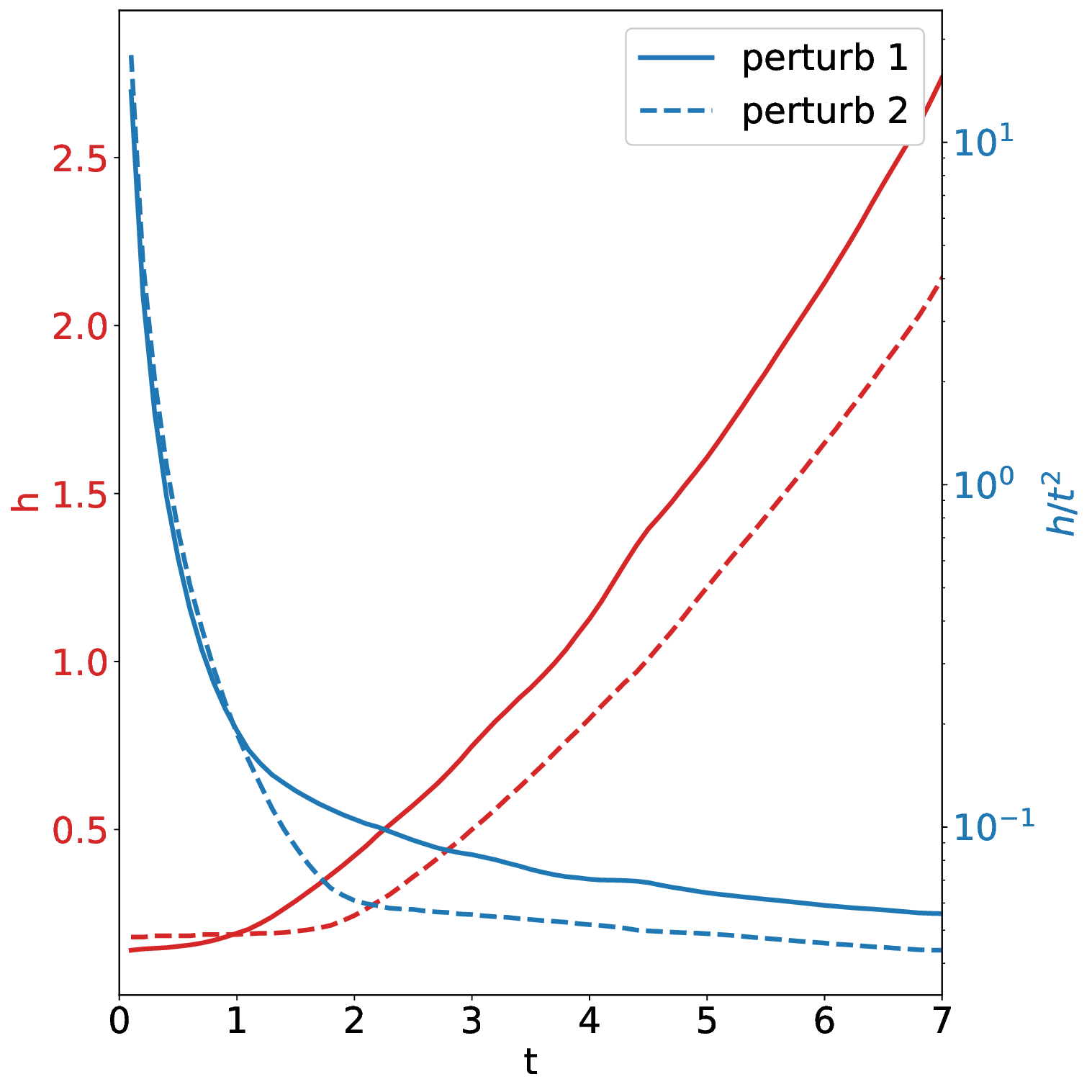}
    \includegraphics[width=0.32\linewidth,height=5cm]{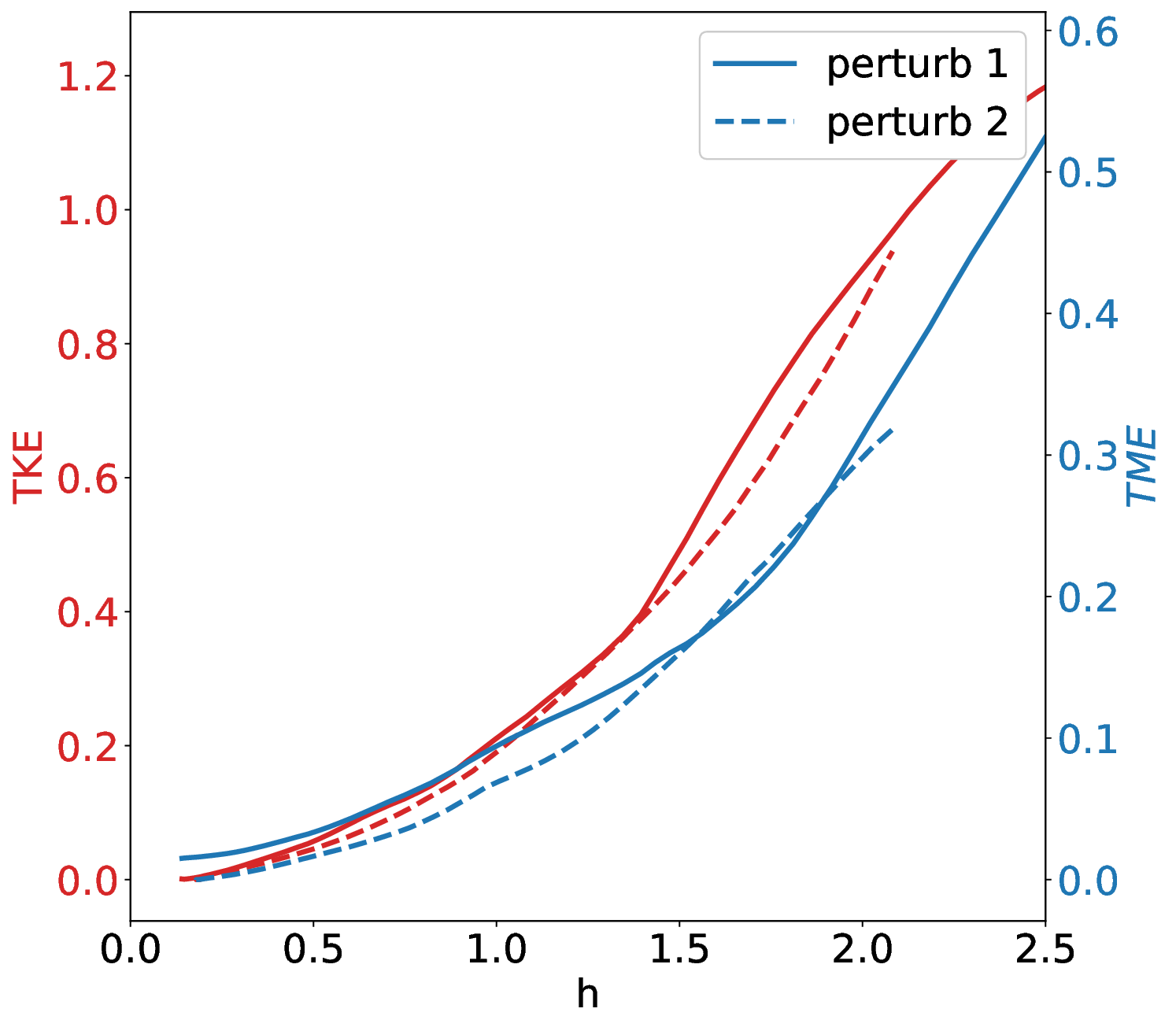}
    \includegraphics[width=0.32\linewidth,height=5cm]{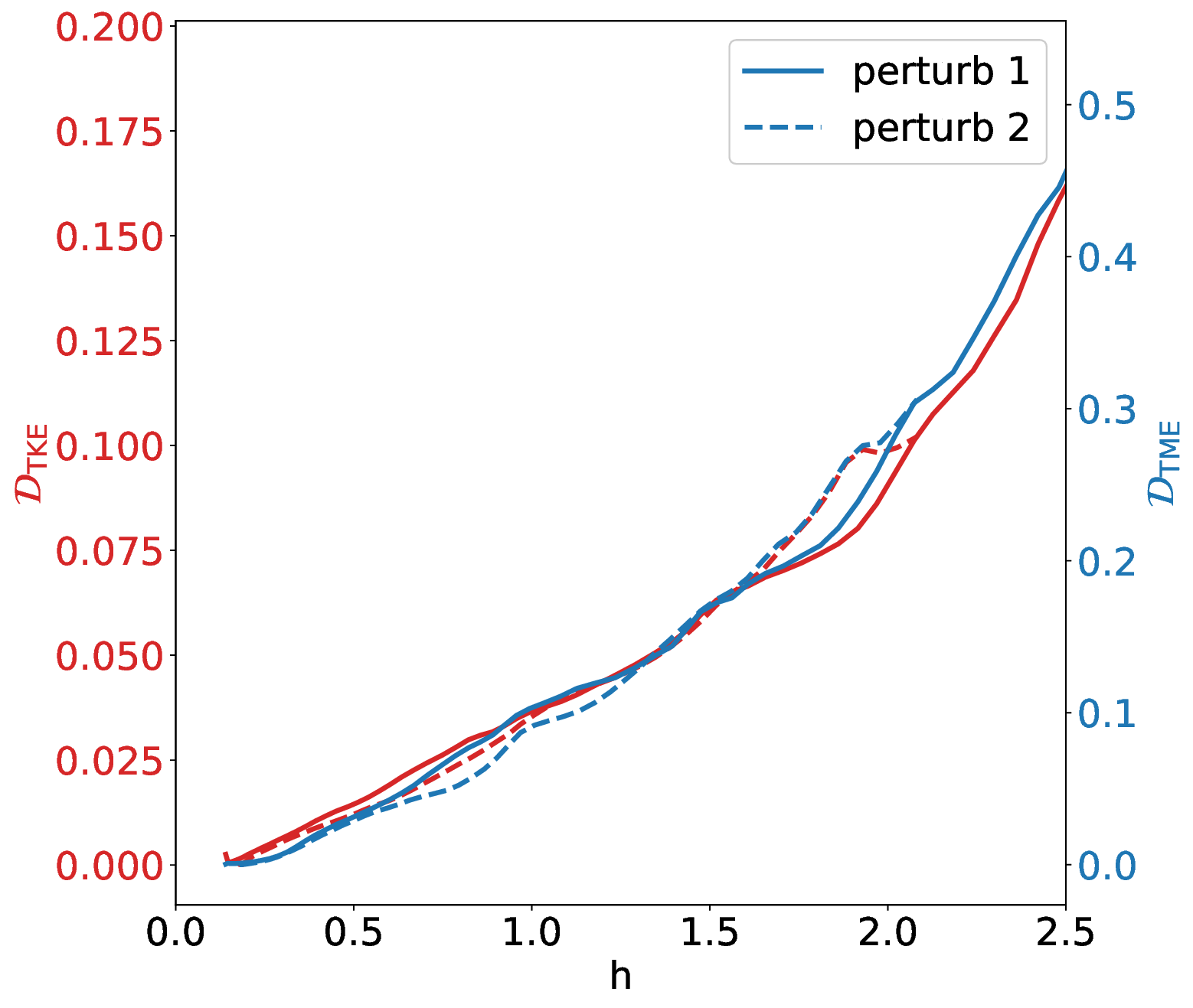}
    \caption{Comparison of 2D MRTI dynamics quantitatively with two perturbations perturb 1 (blue in figure \ref{perturbpower}, solid line here) and perturb 2 (green in figure \ref{perturbpower}, dashed line here). \textit{(left)} Temporal variation of mixing layer height (h); \textit{(center, right)} Variation of TKE, TME, TKE dissipation ($\mathcal{D}_\text{TKE}$) and TME dissipation ($\mathcal{D}_\text{TME}$) with h.}
    \label{perturbcomp}
\end{figure*}

Besides differing in the type of modes present, as discussed in $\S$\ref{Methodology}, the 2D MRTI perturbations are initialized with a higher amplitude and a broader spectrum of wave modes than for 3D. The MRTI evolves faster and we contend that since we compare stages of development by mixing layer height rather than by time elapsed, we do not expect the differences between the initial conditions to have significant effect on the relevant statistics. However, to provide evidence for this, we run a simulation with $k_{max}^{2D} = 90$, matching with $k_{max}$ of the current 3D simulations, $k_{max}^{3D} \left(\equiv\sqrt{(k_x)_{max}^2 + (k_y)_{max}^2} = 90 \right)$. We use smaller perturbation amplitude to make the energy spectrum comparable (see the green and red lines in figure \ref{perturbpower}). The results from this new perturbation (called `perturb 2') are compared against the results from the old perturbation (called `perturb 1') in figure \ref{perturbcomp}. Note that the simulations are compared for a fixed magnetic field strength of $B_0 = 5\% B_c$.

From figure \ref{perturbcomp}(left), we plot the temporal variation of mixing layer height (red, left axis). We note that the low amplitude perturbation (perturb 2) grow much slowly with lower growth rate (blue, right axis) compared to the high amplitude perturbation  (perturb 1). But, if we focus on the dynamical quantities like TKE, TME, and their dissipation, we find that they are similar. In figure \ref{perturbcomp} (centre) we the variation of TKE (red, left axis) and TME (blue, right axis) with mixing layer height. The dynamics are quantitatively and qualitatively similar for both the perturbations. Similarly, from figure \ref{perturbcomp}, the variation TKE dissipation ($\mathcal{D_\text{TKE}}$, red, left axis) and TME dissipation ($\mathcal{D_\text{TME}}$, blue, right axis) with mixing layer height are similar for both the perturbations. Thus we show that the growth rate of the mixing layer height is sensitive to the initial condition, but that the energetics remain comparable and so it is not unreasonable to choose perturbations as described. 

\section{Conclusion}

It is typical for 2D MRTI simulations to be used to understand the dynamics in complex natural systems like accretion discs, supernova, solar prominences \citep{Zhdankin2023, Porth2014, Popescu2021}. But the 2D and 3D are inherently different, with 2D MRTI supporting either undular or interchange modes, but the 3D MRTI supports undular, interchange, and mixed modes. Each of these modes experience the magnetic field differently. This raises an important question: how does the interchange and mixed modes change the instability dynamics? To address this, we compared 2D (with undular modes) and 3D MRTI simulations across three representative magnetic field strengths: (i) a weak field ($B_0 = 1\% B_c$), where MRTI is expected to closely resemble HD RTI; (ii) an intermediate field ($B_0 = 5\% B_c$); and (iii) a strong field ($B_0 = 15\% B_c$), where magnetic effects are prominent.

We find that the interchange and mixed modes have significant influence on MRTI dynamics. For example, owing to less influence from magnetic field these modes create small-scale structures, allow greater fluid mixing (cf. figure \ref{thetaphi}) and stirring with a shorter stirring time scale (cf. figure \ref{djrhodirhodiuj}). The mixing layer of 2D MRTI is characterized by large-scale vortex structures, cf. the higher vorticity length scale (see figure \ref{vorticitylengthscale}). These were established based on diagnostics such as the mixing and interpenetration parameters (equations \ref{mixing_parameter}, \ref{interdisperse}). Due to low mixing and high interpenetration, the released GPE is higher in 2D (cf. figure \ref{GPE}). 

The energy flow in the system begins with energy extraction from GPE by vertical velocity ($\delta \rho g u_z$). A part of this goes into deforming magnetic field lines ($E_{u \rightarrow b}$, see blue line in figure \ref{ubdbquantchap3}). The proportion of energy used to deform the magnetic field lines is lower in 3D MRTI compared to 2D. This is because rising bubbles can partially displace or interchange field-line bundles in the third dimension (where one is available), reducing (relative to a purely two-dimensional domain) the amount of magnetic-line bending required for a given displacement. Some of the magnetic energy subsequently returns to kinetic energy ($E_{b \rightarrow u}$), for example, through reconnection processes. As field strength increases, TKE decreases and TME increases. Additionally, both 2D and 3D systems exhibit energy anisotropy, preferentially in the direction of gravity. This anisotropy is predominant in 3D and intensifies with field strength, cf. figure \ref{aniso}.

Due to small-scale vortical structures, the fraction of energy dissipated is higher in 3D, particularly in the weak field case ($\approx 30\%$ of GPE, compared to $\approx 10\%$ of GPE in 2D, see figure \ref{diss}). The differences in the characteristic length scales (which are influenced by existing wave modes) in the systems, the trend of energy dissipation with magnetic field strength is different in 3D and 2D. The 3D (2D) exhibits a monotonic (non monotonic) trend. The peak dissipation occurs at $B_0 = 5\% B_c$ in 2D, and at $B_0 = 1\% B_c$ in 3D.

The differences between 2D and 3D MRTI extends to the fundamental nature of instability evolution. We find that the 2D MRTI simulations \textbf{may} not be self similar like the 3D case. That is $h \propto t^2$ may be achieved in 3D, but our 2D simulations do not show this behaviour (cf. figures \ref{B0udb_rhov}, \ref{hAgh}). This is due to the presence of interchange and mixed modes in 3D which evolve with no influence or little influence (respectively) from magnetic fields.

Besides revealing the influence of interchange and mixed modes, the above observations conclude that the 2D physics is quite different from 3D in many ways, as summarized in table \ref{table:1}. Hence, we must be cautious while attempting to understand the MRTI physics in the astrophysical systems using 2D simulations. The 2D simulations particularly differ in fluid mixing; energy partition (among TKE, TME, and dissipation), and non-linear instability evolution. However, we find that the conversion rate from magnetic to kinetic energies is very similar in 2D and 3D. This indicates that in the context of magnetic reconnection (a major contributor to magnetic to kinetic energy transfer), 2D simulations \textit{could} be acceptable.

\begin{table}[t]
\centering 
\begin{tabular}{||c | c | c ||} 
 \hline
  & 3D & 2D \\ [0.5ex] 
 \hline\hline
 Dispersion &  & Higher \\ 
 \hline
 Mixing & Higher &  \\
 \hline
 Stirring & Higher &  \\
 \hline
 Released GPE &  & Higher \\
 \hline
 Energy to bend B lines &  & Higher \\ [1ex] 
 \hline
 ME to KE & Similar & Similar \\
 \hline
 Energy dissipation & Higher &  \\
 \hline
 TKE, TME &  & Higher \\ [1ex] 
 \hline
 Energy anisotropy & Higher &  \\ [1ex] 
 \hline
\end{tabular}
\caption{Comparing different phenomena between 2D and 3D. Here, when the difference between 2D and 3D is less than $10\%$ with respect to the 2D or 3D data, we classify it as similar. Higher/lower than the $10\%$ is labelled as higher/lower.}
\label{table:1}
\end{table}

\begin{acknowledgments}
\vspace{-5pt}
MTK is supported by the Engineering and Physical Sciences Research Council (EPSRC) Grant No. EP/W523859/1. AH is supported by STFC Research Grant No. ST/V000659/1, ST/Y00230X/1 and by UKRI Research Grant No. UKRI1165. This work used the University of Exeter High-Performance Computing facility and the DiRAC Memory Intensive service (Cosma7) at Durham University, managed by the Institute for Computational Cosmology on behalf of the STFC DiRAC HPC Facility (www.dirac.ac.uk). The DiRAC service at Durham was funded by BEIS, UKRI and STFC capital funding, Durham University, and STFC operations grants. DiRAC is part of the UKRI Digital Research Infrastructure.
\end{acknowledgments}

\vspace{-10pt}
\section*{Data Availability Statement}
 The data that support the findings of this study are available from the corresponding author upon reasonable request.

\section*{Author contributions}
M.T.K.: data curation, formal analysis, investigation, methodology, software, validation, visualization, writing —original draft, review and editing; A.H.: funding acquisition, project administration, resources, supervision, writing—review and editing

\bibliography{apssamp}

\end{document}